\documentclass[twocolumn,prb,10pt,superscriptaddress]{revtex4-2}
\usepackage[english]{babel}
\usepackage{textcomp}
\usepackage{amsmath}
\usepackage{amssymb}
\usepackage{xcolor}
\usepackage{multirow}
\usepackage{mathtools}
\usepackage{float} 
\usepackage{fontenc}
\usepackage{graphicx}
\usepackage{siunitx}
\usepackage{graphics}
\usepackage{hyperref}
\usepackage{lipsum}
\usepackage{wasysym}
\usepackage{soul}
\usepackage{combelow}
\makeatletter

\def\ket#1{\mathinner{|{#1}\rangle}}

\begin{document}

\title{Stability of destructive quantum interference antiresonances in electron transport through graphene nanostructures}
\author{Angelo Valli}
\affiliation{Institute for Theoretical Physics, Vienna University of Technology, Wiedner Hauptstrasse 8-10, A-1040 Vienna, Austria}
\affiliation{Department of Theoretical Physics, Institute of Physics, Budapest University of Technology and Economics, M\"uegyetem rkp. 3., H-1111 Budapest, Hungary}
\author{Thomas Fabian}
\affiliation{Institute for Theoretical Physics, Vienna University of Technology, Wiedner Hauptstrasse 8-10, A-1040 Vienna, Austria}
\author{Florian Libisch}
\affiliation{Institute for Theoretical Physics, Vienna University of Technology, Wiedner Hauptstrasse 8-10, A-1040 Vienna, Austria}
\author{Robert Stadler}
\affiliation{Institute for Theoretical Physics, Vienna University of Technology, Wiedner Hauptstrasse 8-10, A-1040 Vienna, Austria}

\begin{abstract}
We investigate the stability of destructive quantum interference (DQI) in electron transport 
through graphene nanostructures connected to source and drain electrodes. 
The fingerprint of DQI is the presence of an antiresonance in the transmission function, 
and its origin is deeply connected to the topology of the atomic structure, 
which we discuss in terms of symmetry arguments supported by numerical simulations. 
A systematic analysis of the influence of system size on the transmission
function reveals that the DQI antiresonance persists for large systems in the ballistic regime
and establishes the quantum confinement gap as the intrinsic resolution limit to detect QI effects.  
Furthermore, we consider the influence of disorder, electron-electron and electron-phonon interactions, 
and provide quantitative criteria for the robustness of DQI in their presence. 
We find that the conductance is quite sensitive to perturbations, and its value alone may not be sufficient to characterize the QI properties of a junction. 
Instead, the characteristic behavior of the transmission function is more resilient, 
and we suggest it retains information on the presence of an antiresonance even if DQI is partially concealed or suppressed. 
At the same time, DQI results in a non-linear transport regime in the current-bias characteristics that can be possibly detected in transport experiments. 
\end{abstract}

\maketitle


Quantum interference (QI) in electron transport is a purely quantum mechanical phenomenon 
of keen interest in the field of molecular electronics. 
In single-molecule junctions, experimental evidence of both 
destructive~\cite{guedonNatNano7,frisendaNatChem8,chenCRPS2,chenCCL33,greenwaldNatNano16,baiNatMat18} (DQI) 
and constructive~\cite{vazquezNatNano7,hurtado-gallegoNL22,wangJACS142} (CQI) quantum interference, 
as well as their control~\cite{aradhyaNatNano8,tangJACS143} has been extensively reported.
Sharp antiresonances due to QI with asymmetric Fano shapes and symmetric Mach-Zehnder shapes 
can be found and theoretically explained~\cite{nozakiJPCS427,lambertRSC44,gengJACS137,sangtarashJACS137,sangtarashNanoscale8,hansenJPCC120,manriqueNatComm6}.
Close to the Fermi level, sharp antiresonances drastically affect 
the transport properties~\cite{gehringNL16,sadeghiPNAS112,weiJPLA376,qiuPCCP16},
and indeed, QI has been proposed as a paradigm for a wide spectrum of technological applications, 
ranging from logic gates~\cite{stadlerNanotech15,sangtarashJACS137} single-molecule transistors~\cite{cardamoneNL6,staffordNanotech18},
molecular switches~\cite{greenwaldNatNano16,daaoubNanomat10}, spin-filters~\cite{lundebergNatPhys5,valliNL18,valliPRB100,palNatComm10},
as well as for enhancing the performance of thermoelectric~\cite{chenCCL33,caoJPCM31,almughathawiACSS6,hurtado-gallegoNL22} 
and chemical sensing~\cite{weiJPLA376,prasongkitRSCA6,sengulPRB105} devices.  

Fundamentally, QI effects occur when the transmission of electrons across a resistor is phase-coherent,  
which is realized when the length of the transmission channel $ \cal L$ is shorter than the characteristic scale (mean free path)  
associated with elastic ($\lambda_e$) and phase-breaking ($\lambda_\phi$) electron scattering events, i.e., ${\cal L} < \lambda_{e}, \lambda_{\phi}$ (ballistic regime). 
For large ballistic cavities of size $L$, an intuitive interpretation is provided by
semiclassical approaches like Gutzwiller's trace formula \cite{BrackBhaduri}. By contrast, QI in the context of molecular electronics rather originates from contributions between different molecular orbitals (MO), 
which couple to the electrodes with similar strength but different phases~\cite{zhaoJCP146}. 
A qualitative understanding of QI can be achieved within a single-particle picture
making QI deceptively simple. 
However, a careful treatment is warranted, since the loss of electronic phase coherence is typically (but not exclusively) associated to inelastic processes 
which are often neglected in theoretical calculations, 
such as, e.g., electron-electron or electron-phonon scattering. 

Notwithstanding different sources of scattering, which can possibly spoil phase coherence, 
QI effects can be detected under experimental conditions. 
While early evidence of QI has been rather indirect~\cite{fracassoJACS133},
the dramatic technological progress of the last few decades eventually lead to a clear direct detection of DQI~\cite{guedonNatNano7}.

It is worth stressing that current fabrication techniques do not offer high enough spatial resolution to reliably contact single molecules with well-defined geometries. Instead, transport measurements of single-molecule devices 
are typically performed with break-junction techniques, either in a mechanically-controlled 
or an scanning tunneling microscope setup~\cite{eversRMP92}.
In the simplest case, a thin gold wire is mechanically strained until the wire breaks, creating two fresh electrode surfaces that can now be connected by individual molecules in a self-assembled way. Consequently, the junction is subject to configuration fluctuations, as 
the atomic arrangement is unknown. 
The measurement is therefore repeated 
over a series of break-junction events to obtain reliable statistics. 
Despite the statistical nature of break-junction experiments, 
there are protocols to identify the suppression of the electronic transmission due to DQI, e.g., 
in differential conductance ($dI/dV_b$) maps~\cite{guedonNatNano7,zhangCS9,bessisSR6}, 
or through the analysis of conductance histograms~\cite{baiNatMat18,tangJACS143,wangChemComm57,frisendaNatChem8,arroyoNRL8} 
of single-molecule junctions. 
Remarkably, QI effects seem also to be surprisingly stable, 
having been detected even at room temperature~\cite{guedonNatNano7,aradhyaNatNano8,arroyoNRL8,wangJACS142},
and on length scales well beyond that of single-molecule junctions, 
in systems such as, e.g., macromolecules~\cite{richertNatComm8} as well as nanostructured graphene~\cite{valliNL18,valliPRB100,calogeroJACS141,canevaNatNano13,gehringNL16}.

In particular, graphene represents a natural platform for high performance nanoelectronics~\cite{satoJJAP54}, 
thanks to its unique physical properties, 
and its ability to form atomically precise nanostructures~\cite{caiNat466,guimaraesACSNano10,wangSci342,rhodesNatMat18,canevaNatNano13}.
Antiresonances akin to those occurring in single-molecule junctions 
have been predicted theoretically~\cite{liPRB77,yinJAP107,wakabayashiPRL84,darancetPRL102,gunlyckeAPL93,valliNL18,valliPRB100,nitaPSSRRL8,nitaPRB101} 
and confirmed experimentally~\cite{gehringNL16,rutterSci317,yangNL10,oksanenPRB89,canevaNatNano13}. 
QI effects can also occur in the diffusive regime, i.e., $\lambda_e \ll {\cal L} < \lambda_{\phi}$, 
where weak localization can arise due to the coherent superposition of random scattering paths~\cite{bischoffPRB90}. 
Since the effective mean free path of electrons in graphene strongly depends on the local doping~\cite{giannazzoNL11}, 
both regimes may be relevant for a single device at different energies. 

In the present work we analyze, within a unified framework, the stability of QI antiresonances in graphene nanostructures. 
We show that the the ballistic transmission function displays a characteristic behavior within the quantum confinement gap, 
that can be entirely ascribed to the existence of an antiresonance. 
We consider a wide range of different mechanisms which are detrimental to QI, 
including disorder, electron-electron, and electron-phonon interactions and quantify their effect on a QI antiresonance. Our results thus provide stringent boundaries on the possibility of detecting QI effects in an experimentally-relevant parameter range.

\section{Topological conditions for DQI}\label{sec:dqi}

Predicting the occurrence of QI in the electron transmission function  
is a challenging task~\cite{stadlerNanotech15,markussenNL10,markussenPCCP13,pedersenNL15,stadlerNL15,nozakiJPCS427}. 
A few \emph{back-of-the-envelope} methods have been developed~\cite{stadlerNanotech15,markussenNL10,pedersenNL15,stuyverJPCC119,odriscollNanoscale13,gengJACS137}, 
including a graphical scheme, which relies on a visual inspection of the molecular structure 
and the topology of the H\"uckel (or tight-binding) Hamiltonian,
and is able to predict DQI without the need for explicit numerical simulations~\cite{stadlerNanotech15}. 
This graphical scheme has been validated against density functional theory~\cite{markussenNL10}, 
it was extended to hetero-atoms~\cite{markussenPCCP13}, 
non-alternant hydrocarbons such as azulene~\cite{pedersenNL15,stadlerNL15}, 
and further generalized in a diagrammatic fashion 
to calculate the position of the antiresonances~\cite{pedersenNL15}. 
The Coulson-Rushbrooke pairing theorem from quantum chemistry~\cite{coulsonMPCPS36} has
recently reconciled the graphical approach with a MOs perspective more common for 
a theoretical analysis of molecular properties~\cite{zhaoJCP146}.

The graphical scheme and the pairing theorem naturally link the sublattice structure to DQI. 
The term alternant hydrocarbons refers to conjugated hydrocarbon systems where carbon atoms can be divided into two subsets 
(or sublattices) with nearest-neighbor interactions only between two atoms of the two different subsets -- i.e., any hydrocarbon
system containing no odd-membered rings.
For such systems a QI antiresonance appears between the highest occupied (HOMO) and the lowest unoccupied (LUMO) MOs~\cite{zhaoJCP146,tsujiJCP149,tsujiCR118,valliNL18,valliPRB100,nitaPRB101} if the contact sites belong to the same sublattice. 
If the contact sites belong to different sublattices, DQI is still possible if certain conditions are met~\cite{nitaPRB101}, 
but it is in general harder to predict. 
These properties have also been confirmed within the more general framework 
of Green's functions formalism,~\cite{valliNL18,valliPRB100,nitaPSSRRL8,nitaPRB101,nitaPRB103,pedersenPRB90}
which also revealed a rich QI phenomenology~\cite{pedersenPRB90}. 
Since graphene nanostructures can be considered, and even chemically synthesized~\cite{naritaNatChem6,caiNat466}, 
as extended polyaromatic molecules with an alternant structure and hydrogen-passivated edges, the sublattice scenario naturally holds. 
DQI has been demonstrated numerically for rectangular~\cite{nitaPSSRRL8,nitaPRB101} and hexagonal~\cite{valliNL18,valliPRB100} graphene nanostrutures.
Indeed, such an argument predicts DQI in graphene nanostructure 
with any size and shape, with some \emph{caveats}~\cite{nitaPRB101,nitaPRB103}. 
Such a topological argument greatly simplifies the analysis, 
as it provides a very simple criterion to identify 
which contact configurations display a QI antiresonance. 
It is equally clear, however, that for larger graphene flakes, more phases accumulate, and
as a consequence, the characteristic dip due to DQI should become narrower. Indeed, only the existence of the dip is topologically protected - there is no general statement
on its width. Additional interactions such as electron-electron or electron-phonon coupling will further affect the phase coherence required for quantum
interference effects. Consequently, one might ask under which realistic conditions DQI effects can still be
measured. In the following, we intend to provide a comprehensive answer to this question, investigating
the role of system size, disorder, electron-phonon, and electron-electron coupling on the DQI dip.

\begin{figure*}[thp]
\includegraphics[width=\linewidth, angle=0]{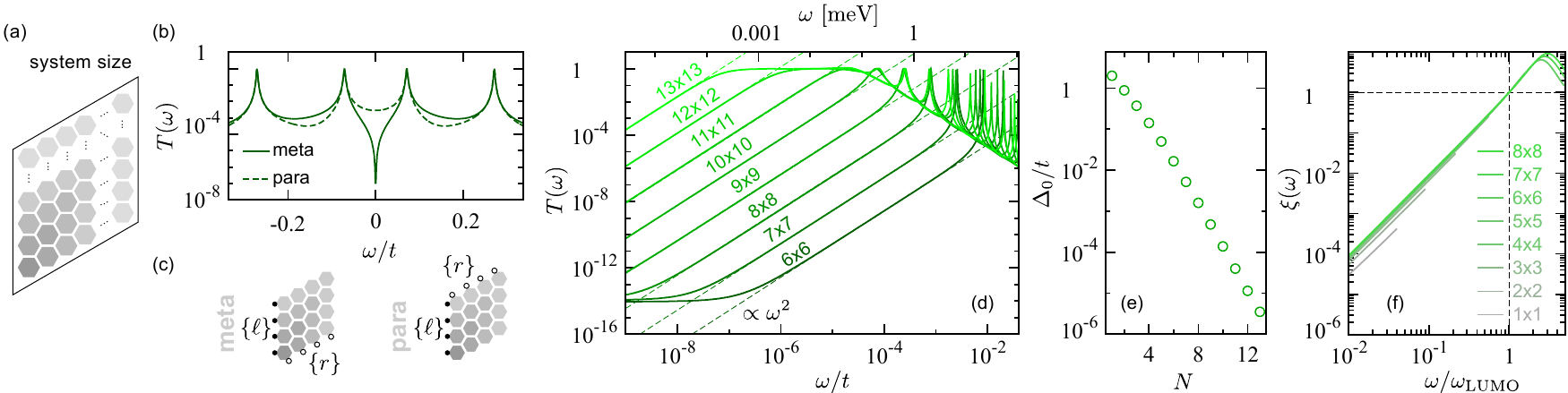}
\caption{Effects of system size on DQI. (a) Representative $N \times N$ graphene nanostructure.   
(b) Transmission function in the meta an para configurations for a $4 \times 4$ nanostructure. 
(c) Connections in the meta and para configurations, the black and white circles indicate the sets of AO $\{\ell\}$ and $\{r\}$ connected to the L and R leads, respectively. 
(d) Evolution of the transmission function in the meta configuration with system size. 
Within the gap $T(\omega) \propto \omega^2$ and saturates as $\omega \rightarrow \omega^{\mathrm{DQI}}$. 
(e) Evolution of the HOMO-LUMO gap $\Delta_0$ with system size. 
(f) Data collapse of the ratio ${\xi}$ between the transmission function in the meta and para configurations, for system size $N \gtrsim 4$. }
\label{fig:effect_size}  
\end{figure*}

\section{Results}

We consider a graphene nanostructures with rhomboidal shape and size $N\!\times\!N$, 
where $N$ denotes the number of rings along each edge, and $n_C$ the number of C atoms in the nanostructure
[see Fig.~\ref{fig:effect_size}(a)].
As we discuss below, our arguments are rather general 
and are expected to remain valid for graphene nanostructure with other shapes (e.g., rectangular or hexagonal). 

We begin our analysis by studying the dependence of DQI features on system size within the tight-binding approximation. 
The corresponding Hamiltonian, describing a single $p_z$ AO per C atom, reads
\begin{equation}\label{eq:tight-binding}
 {\cal H}_0 = t \sum_{\langle ij \rangle} \sum_{ \sigma} c^{\dagger}_{i\sigma} c^{\phantom{\dagger}}_{j\sigma}, 
\end{equation}
where $c^{(\dagger)}_{i\sigma}$ is the annihilation (creation) operator 
for an electron on site $i$ with spin $\sigma$, 
and $t$ is the hopping between nearest-neighbor sites $i$ and $j$, 
for which we take a typical value $t = \SI{2.7}{\eV}$. 
Including longer-range hoppings, or deriving them \textit{ab-initio}~\cite{gandusJCP153}, 
e.g., breaks the particle-hole symmetry and shifts the position 
of the antiresonance within the gap~\cite{valliNL18} 
but does not invalidate the conclusions of our analysis.  

Since the tight-binding Hamiltonian ${\cal H}_0$ does not include any scattering mechanism (neither elastic nor inelastic), 
in this approximation the electronic excitations do not decay, i.e., are characterized by a lifetime $\tau = \infty$ and mean-free paths $\lambda_e = \lambda_{\phi} = \infty$. 
As we introduce disorder, electron-electron, or electron-phonon scattering below, 
we estimate the characteristic length scales on which QI effects are suppressed. 

\subsection{Universal transport behavior due to DQI} 

We describe electronic transport within the Landauer formalism~\cite{landauerRD1}, where the transmission function is given by 
\begin{equation} \label{eq:landauer}
 T(\omega)=\mathrm{Tr}\Big[ \Gamma^{L}(\omega)G^{\dagger}(\omega)\Gamma^{R}(\omega)G(\omega)\Big], 
\end{equation}
in terms of the molecular Green's function $G(\omega)$ and the coupling matrix $\Gamma^{L/R}$ to the left ($L$) and right ($R$) leads.
Let us further denote by $\{\ell\}$ and $\{r\}$ the sets of edge AOs contacted to the $L$ and $R$ leads, respectively. 
Within a wideband limit (WBL) approximation for the leads, the diagonal coupling elements 
$\Gamma^{L}_{\ell\ell}$ and $\Gamma^{R}_{rr}$ are given by energy-independent constants $\Gamma$ 
for edge C atoms $\{\ell\}$ and $\{r\}$, and zero otherwise (see also SI). 
In the following, we set all non-zero couplings to the leads to $\Gamma=0.0004t \approx \SI{1}{\milli\eV}$ 
(or  $\approx \SI{10}{\kelvin}$, for reference) unless otherwise specified. 
We have verified numerically that neither the WBL approximation~\cite{valliNL18,verzijlJCP138} 
nor neglecting non-diagonal couplings $\Gamma_{i\neq j}$~\cite{reuterJCP141,tsujiJCP141,hansenJPCC120,samangNJP19}  fundamentally change the QI properties of the system under study. 

We decompose the transmission function as a sum over independent channels 
\begin{equation} \label{eq:Te_channels}
 T(\omega) = \sum_{\ell, r} T_{\ell \rightarrow r}(\omega) = \Gamma^2 \sum_{\ell r} |G_{\ell r}(\omega)|^2.
\end{equation}
In the non-resonant transport regime, 
the condition for an antiresonance in the transmission function due to DQI becomes $\Re G_{\ell r}(\omega_{\mathrm{DQI}})=0$, 
for a given frequency $\omega_{\mathrm{DQI}}$~\cite{valliNL18,valliPRB100,pedersenPRB90}. 
Indeed, we find a pronounced antiresonance at $\omega=0$ in a $4\times 4$ graphene
nanostructure (solid line in Fig.~\ref{fig:effect_size}b) 
if $\ell$ and $r$ belong to the same sublattice 
(labeled as \emph{meta} configuration, Fig.~\ref{fig:effect_size}c). 
If $\ell$ and $r$ instead belong to different sublattices (labeled as \emph{para} configuration, Fig.~\ref{fig:effect_size}d)
there is, in general, no antiresonance, and the transmission function within the HOMO-LUMO gap saturates at a much higher value 
[dashed line in Fig.~\ref{fig:effect_size}(b)]. 
These results are consistent with the topological conditions for DQI discussed above. 


In the meta configuration, we identify a \emph{universal} behavior of the transmission function at $\omega=0$. 
We calculate the transmission function $T(\omega>0)$ 
through graphene nanostructures of increasing size (Fig.~\ref{fig:effect_size}(d))
and find a dip only in the meta configuration.
(the transmission function for $\omega<0$ remains symmetric due to the particle-hole symmetry in our model). 
The evolution of the LUMO resonance at $\omega_{\mathrm{LUMO}}$ shows that 
the HOMO-LUMO gap, defined as $\Delta_0=\omega_{\mathrm{LUMO}}-\omega_{\mathrm{HOMO}}$, 
decreases with systems size (Fig.~\ref{fig:effect_size}(e)), 
in agreement with the literature~\cite{sonPRL97,gucluPRB82,singhJCP140}. 
The transmission function saturates close to the antiresonance $\omega \rightarrow \omega_{\mathrm{DQI}}$ 
(which is \emph{pinned} at $\omega_{\mathrm{DQI}}=0$ due to particle-hole symmetry)
while at higher energies $|\omega| \lesssim |\omega_{\mathrm{LUMO}}-\Gamma|$, it follows the universal behavior
\begin{equation}\label{eq:tomega}
 T_{\mathrm{metha}}(\omega) = \alpha_N \cdot 10^N \cdot (\omega/t)^2,
\end{equation}
with $\alpha_N$ a size-dependent constant. 
Such a behavior follows naturally from Eq.~(\ref{eq:Te_channels}) 
as $\Re G_{\ell r}(\omega \rightarrow \omega_{\mathrm{DQI}}) \propto \omega$~\cite{markussenPRB89}, 
yet it extends over the whole energy gap. 

From the numerical data we observe that the saturation value $T(\omega_{\mathrm{DQI}})$ is weakly size-dependent. 
However, at a finite energy scale, e.g., due to experimental resolution, 
the transmission increases approximately by an order or magnitude when increasing $N$ by one unit. 
This is because $\Delta_0$ decreases with size but the coupling between molecule and leads $\Gamma$, 
which determines the width of the transmission features, remains constant. 
An analogous dependence of the transmission function is observed at constant $\Delta_0$ by increasing $\Gamma$ (see SI). 
To remove this contribution, we consider the ratio ${ \xi}(\omega) = T_{\mathrm{meta}}(\omega) / T_{\mathrm{para}} (\omega)$ 
between the transmission in the meta and para configurations [Fig.~\ref{fig:effect_size}(f)]. 
Away from the Dirac point, i.e., for $\left|\omega/t\right| \gg 0$ the transmission
in meta and para configurations approach each other, ${\xi}(1) \rightarrow 1$ 
because the position and the width of the LUMO and HOMO resonances are (approximately) the same in the meta and para configurations, 
cfr. Fig.~\ref{fig:effect_size}(b). 
For energies within the HOMO-LUMO gap, all the curves $\xi(\omega)$ \emph{collapse} onto each other for $N \ge 4$ [see Fig.~\ref{fig:effect_size}(f)].
The underlying reason is related to the distribution of poles of the transmission function on the complex plane: for frequencies sufficiently close to zero, i.e., sufficiently far away from the other poles at finite $\omega$, the behavior around the gap becomes universal.
Hence, by disentangling the behavior induced by the antiresonance from $\Delta_0$ and $\Gamma$, 
we can draw the following conclusions: 
(i) for $N \ge 4$, the transmission $T_{\mathrm{meta}}$ within the HOMO-LUMO gap 
approaches the universal form of Eq.~(\ref{eq:tomega}), with the prefactor $\alpha_N$ determined by $T_{\mathrm{meta}}(\omega_{\mathrm{LUMO}}) = T_{\mathrm{para}}(\omega_{\mathrm{LUMO}})$ (i.e., $\xi(\omega_{\mathrm{LUMO}})\rightarrow 1$), and 
(ii) our argument remains valid for differently shaped graphene nanostructures, especially as the surface-to-bulk ratio decreases. 
Hence, in the following we shall focus on $4\times 4$ graphene nanostructures, 
which strike a good balance between numerical cost while displaying 
weak-to-none finite-size effects on the universal behavior of the transmission of Eq.~(\ref{eq:tomega}).

\begin{figure*}[htp]
\includegraphics[width=\linewidth, angle=0]{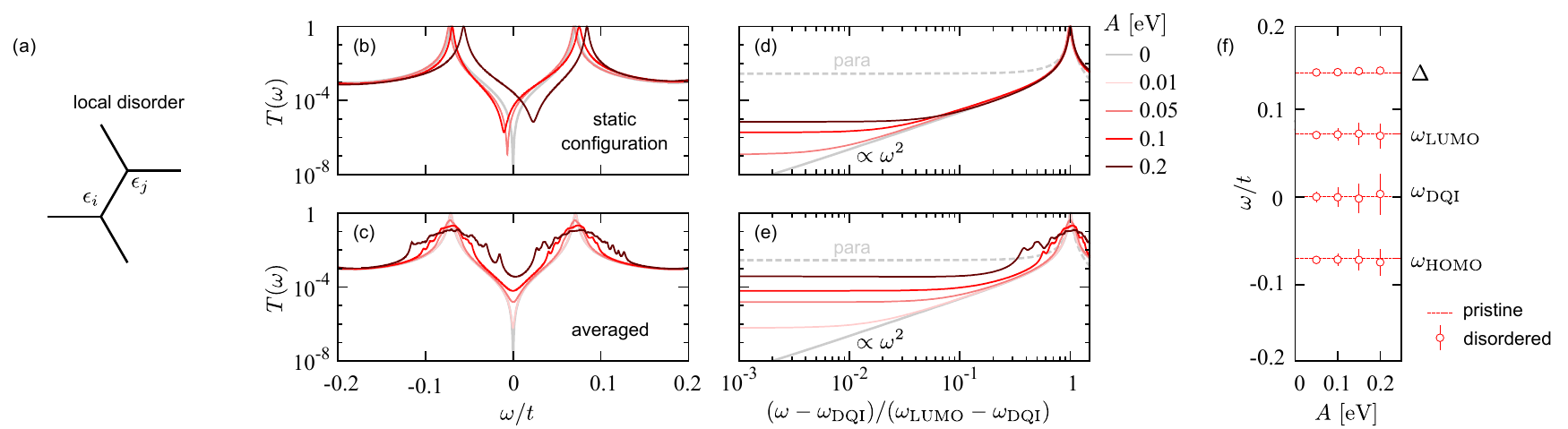}
\caption{(a) Schematic representation of the local disorder. 
(b,c) Transmission function for a specific disorder realization (b) and averaged over hundreds of realizations (c),  
for different values of disorder strength. 
(d,e) Rescaling the energy allows to align the position of the antiresonance $\omega_{\mathrm{DQI}}$ 
and of the LUMO resonance $\omega_{\mathrm{LUMO}}$ for each disorder realization (or their average). 
Disorder strongly enhances the conductance, but the characteristic $\omega^2$ behavior 
is clearly observed over an extended energy range up to moderate disorder strength. 
Dashed grey lines mark the transmission functions of the corresponding pristine para configurations as a reference. 
(f) In contrast to the other transmission features whose variance increases with the disorder strength, a statistical analysis reveals that the size of the HOMO-LUMO gap is remarkably stable 
around the pristine values (red dashed lines). }
\label{fig:effect_disorder}  
\end{figure*}


Remarkably, the universal behavior of the transmission function reflects on the current-bias ($I-V_b$) characteristics, which can be directly measured in the experiment, whereas the transmission 
is loosely related to the differential conductance $dI/dV_b$, 
at least at low bias voltages $eV_b = \mu_L-\mu_R$, where $\mu_L$ ($\mu_R$) is the chemical potential in the left (right) contact. 
We model the bias dependence in the Landauer-B\"uttiker framework, i.e., 
\begin{equation}
 I \approx \frac{e}{h}\int_{\mu_R}^{\mu_L} T(\omega) \mathrm{d}\omega,
\end{equation} 
where $e$ denotes the electric charge and $h$ the Planck constant. 
Within the HOMO-LUMO gap, we can insert the universal behavior of the transmission function. For the meta configuration, the $\omega^2$ behavior of Eq.~(\ref{eq:tomega}) yields a non-linear characteristics $I_{\mathrm{meta}}  \propto V_b + t^{-2} V_b^3$,  
whereas in the para configuration the slowly-varying transmission $T_{\mathrm{para}}(\omega)\approx$ const
results in a linear characteristics $I_{\mathrm{para}} \propto V_b$ (see SI). 
Hence, we identify this specific non-linear transport regime as a fingerprint of DQI, 
which is useful for the experimental characterization of a junction, 
besides the value of the zero-bias conductance $G=(e^2/h) T(0)$.  
In single-molecule junctions, with a gap in the range of a few \SI{}{eV}, 
the experimental resolution necessary to resolve the in-gap $I-V_b$ characteristics is hardly an issue.
By contrast, the HOMO-LUMO gap quickly becomes unmeasurably small for large graphene nanostructures, i.e., below 1 meV for $N = 11$. 
Another possible limitation is the contribution to the transmission function through the $\sigma$ channel, 
which is neglected in our numerical simulation, and can in principle mask interference effects in the $\pi$ channel. 
However, the $\sigma$ contribution is naturally suppressed for longer molecules due to a faster decay with length 
(see, e.g.,~\cite{garnerJCPL11} and references therein) 
and can therefore be expected to be less relevant upon increasing systems size.  

In summary, for larger flakes, the main restriction for observing DQI effects is the energy scale of the HOMO-LUMO gap, which approaches the limit of experimental resolution at $ N \gtrsim 12$.



\subsection{Disorder}\label{sec:disorder}

For exploiting DQI in practice, robustness with respect to moderate disorder is critical. 
In state-of-the-art hBN-graphene sandwich devices, 
bulk disorder is of the order of a few \SI{}{\milli\eV}~\cite{rhodesNatMat18},
with a major contribution from long-range strain modulations~\cite{coutoPRX4}, 
and evidence of ballistic transport exceeding \SI{1}{\micro\meter} has been reported~\cite{banszerusNL16}. 

We investigate the influence of uncorrelated local random disorder 
focusing on the configurations exhibiting DQI. 
The scope of our analysis is two-fold: 
(i) we show how the transmission function changes with respect to the pristine sample, 
for configurations with \emph{static} disorder, and
(ii) we look at the transmission function averaged 
over hundreds of disorder realizations 
(also referred to as \emph{dynamic} disorder)  
by adding the individual transmission traces incoherently. 
The disorder average is representative of the statistical nature of experimental measurements in a break-junction setup.  
The combination of the two effects allows us to understand 
the stability of DQI against disorder. 
We find that QI is surprisingly robust up to disorder amplitudes of $\approx \SI{40}{\milli\eV}$, 
which is at least one order of magnitude above the experimental estimates for state-of-the-art devices. 

The local disorder is described by adding to the tight-binding Hamiltonian (\ref{eq:tight-binding}) the following term
\begin{equation}\label{eq:disorder}
 {\cal H}_{\mathrm{disorder}} = \sum_{i\sigma} \epsilon_i n_{i\sigma}, 
\end{equation}
where $n_{i\sigma} = c^{\dagger}_{i\sigma}c^{\phantom{\dagger}}_{i\sigma}$ is the number operator 
and $\epsilon_{i}$ is the on-site energy of site $i$. We create
random disorder with $\langle\epsilon_i\rangle = 0$ and $\langle\epsilon_i^2\rangle = A^2$, with
$A$ the disorder amplitude.

We calculate the transmission function for different disorder realizations (Fig.~\ref{fig:effect_disorder}). 
Since disorder breaks the particle-hole symmetry, the energies of the MOs, 
and their projection onto the C-$p_z$ AOs no longer fulfill the Coulson-Rushbrooke pairing theorem. 
In particular, the position of the frontier MOs  
($\omega_{\mathrm{HOMO}}$ and $\omega_{\mathrm{LUMO}}$) 
now depends on the specific disorder realization, shifting the characteristic DQI dip  in the transmission function
randomly (Fig.~\ref{fig:effect_disorderb}). The antiresonance, which emerges from the cancellation of contributions involving all MOs~\cite{zhaoJCP146}, 
is no longer pinned at the Fermi energy, i.e., $\omega_{\mathrm{DQI}} \neq 0$. 
Moreover, the cancellation of the transmission is partial, as $\omega_{\mathrm{DQI}}$ becomes channel-dependent (see SI).
Averaging over disorder smears out the signatures of DQI Fig.~\ref{fig:effect_disorder}(c).  
Close to the DQI resonance, the conductance is thus effectively \emph{enhanced} by disorder, both in individual realizations and on average. 
In this sense, we can talk about \emph{disorder-assisted} transport, 
as the role of the disorder is to suppress the QI effects responsible for the transmission minimum. 
Finally, we note that for individual realizations the resonant transport through the MOs remains unitary, 
while the average over disorder introduces an effective decoherence and drives the system away from the ballistic regime.

In order to understand how disorder affects the universal behavior of the transmission function, 
we rescale the conductance traces as a function of the dimensionless scale
$(\omega-\omega_{\mathrm{DQI}})/(\omega_{\mathrm{LUMO}}-\omega_{\mathrm{DQI}})$, 
using the corresponding value of 
$\omega_{\mathrm{DQI}}$ and $\omega_{\mathrm{LUMO}}$ 
extracted for each individual disorder realization [Fig.~\ref{fig:effect_disorder}(d)] 
or their average [Fig.~\ref{fig:effect_disorder}(e)]. 
This aligns the position of both the antiresonance and the gap edge 
for both static and dynamic disorder.    
Focusing on the transmission function close to the antiresonance, 
we find quadratic enhancement with disorder strength, 
i.e., $T(\omega_{\mathrm{DQI}}) \propto A^2$ for both individual configurations and on average (see SI). 
Hence, for any given value of the disorder strength, the transmission function interpolates 
between a constant regime at $\omega \approx \omega_{\mathrm{DQI}}$ and the characteristic $\omega^2$ behavior at higher energies, 
and the crossover scale increases with disorder strength. 
Above a critical threshold, the transmission function becomes qualitatively indistinguishable 
from that of the para configuration. Despite a lower conductance the effects of DQI are lost. 
This scenario emerges for both individual disorder configurations and on average. 
This suggests that an analysis of the conductance alone may not be conclusive for detecting DQI 
while the effects on the transmission function at finite energy 
and the corresponding non-linearity of the $I-V_b$ characteristics are more resilient to the effects of disorder (see SI).  

Disorder may also be characterized by  a disorder scattering length.  
For any given realization of static disorder the electron transport is ballistic   
and the transmission in the resonant regime (e.g., at the LUMO resonance) is unitary,   
hence $\lambda_{\mathrm{disorder}}=\infty$. 
After averaging over the disorder realizations the resonant transmission is reduced as~\cite{datta2005}
\begin{equation}
 T(\omega_{\mathrm{LUMO}}) =  \frac{\lambda_{\mathrm{disorder}}}{{\cal L} + \lambda_{\mathrm{disorder}}}. 
\end{equation}
For $\lambda_{\mathrm{disorder}} \gg {\cal L}$ unitary transport 
is restored.  
Taking as system size ${\cal L} \approx \SI{1}{\nano\metre}$, estimated as the longest distance among all $\ell \rightarrow r$ channels, 
we find, e.g., $\lambda_{\mathrm{disorder}} \approx \SI{1.2}{\nano\metre}$ at $A=\SI{30}{\milli\eV}$, 
and a relation $\lambda_{\mathrm{disorder}}(A) \propto A^{-1}$ (see SI).  
Alternatively, if we express the condition above as 
\begin{equation}
 \frac{\lambda_{\mathrm{disorder}}}{{\cal L}} = \frac{T(\omega_{\mathrm{LUMO}})}{1-T(\omega_{\mathrm{LUMO}})},
\end{equation}
it is possible to estimate a disorder threshold (independent on ${\cal L}$) defined by $\lambda_{\mathrm{disorder}} \approx {\cal L}$, 
corresponding to a resonant transmission reduced to half its ballistic value.  
We find $A_{\mathrm{critical}} \lesssim \SI{40}{\milli\eV}$, 
and verified that when approaching the critical disorder strength 
the $\omega^2$ transmission disappears, 
meaning that the effects of QI are contextually lost (see SI). 
Experimentally, the critical value is comparable with disorder estimates for graphene on a SiO$_2$ substrate
but well above estimates for devices encapsulated in hBN~\cite{rhodesNatMat18}. 
For individual disorder realizations, QI are lost only at much higher values of disorder strength (i.e., above those investigated here).
We conclude that the intrinsic (i.e., static) disorder of the sample is not as detrimental to QI as averaging over several disorder configurations, 
which can be considered representative of a statistical analysis of the transport properties over a series of break-junction configurations. 

Finally, we  compare the disorder average and variance to a few energy scales relevant for electron transport, 
i.e., the position of the resonant ($\omega_{\mathrm{HOMO}}$ and $\omega_{\mathrm{LUMO}}$) and interference ($\omega_{\mathrm{DQI}}$) features, and the gap [Fig.~\ref{fig:effect_disorder}(f)].
Since the disorder distribution is symmetric, i.e., $\langle \epsilon_k \rangle = 0$, 
the mean value of all quantities is close to that of the pristine sample, 
while their variance generally increases with disorder strength. 
As already discussed, the fluctuation of $\omega_{\mathrm{DQI}}$ in the individual transmission channels for each disorder realization 
is responsible for the enhancement of the conductance, see Figs.~\ref{fig:effect_disorder}(b,c,d,e).
However, we also find that the gap $\Delta$ is remarkably stable against disorder, 
i.e., its variance is significantly lower than the variance of the frontier MOs position 
for any value of the disorder strength. 
This suggests that the dominant effect at play is rather a fluctuation of the Fermi level alignment. 
The data are also compatible with a weak \emph{increase} of the gap, 
e.g., less than $2\%$ with respect to the pristine value $\Delta_0$ at $A \approx \Delta_0$. 

Overall, limiting disorder to below 10 meV --- which is readily possible for state-of-the-art substrates like hexagonal boron nitride \cite{banszerusNL16} ---  sufficiently limits disorder effects to still be able to observe DQI.

\subsection{Electron-electron interaction}

Electronic correlation arising from the Coulomb repulsion are believed to play a relevant role in electron transport, 
when electrons are constrained in narrow conduction channels. 
Unfortunately, taking into account many-body effects in the theoretical description of electron transport is a challenging task~\cite{eversRMP92},  
and despite attempts to include dynamical correlations (in different fashions) in the recent literature~\cite{thygesenPRB77,strangePRB83,markussenPRB89,jacobJPCM27,valliNL18,valliPRB100,kropfPRB100,droghettiPRB105,droghettiPRB106},
systematic studies for correlated nanoscale quantum junctions are still few and far between. 

In the following, we add to the tight-binding Hamiltonian a screened local Coulomb repulsion (i.e., a Hubbard interaction) described by the Hamiltonian 
\begin{equation}\label{eq:Hubbard}
 {\cal H}_{\mathrm{e-e}} = U \sum_i n_{i\uparrow} n_{i\downarrow} - \mu \sum_i (n_{i\uparrow} + n_{i\downarrow}),
\end{equation}
where the chemical potential to be $\mu=U/2$ ensures the the Fermi energy is located at $\omega=0$ for any value of the Coulomb repulsion.  
We take into account the many-body effects due to electron-electron interaction within real-space dynamical mean-field theory (DMFT). 
In a nutshell, each C atom is mapped onto an auxiliary impurity problem (describing a single C-$pz$ AO) self-consistently embedded in the nanostructure,  
which is solved self-consistently with L\'{a}nczos exact diagonalization,~\cite{weberPRB86,amaricciCPC273} 
similarly as in previous works~\cite{valliPRB94,valliNL18,valliPRB100,baumannPRA101}. 
All the many-body effects are enclosed in the electronic self-energy $\Sigma_{ij}(\omega)=\Sigma_{ii}(\omega)\delta_{ij}$, 
which is a diagonal matrix in the space of the carbon $p_z$ AOs.  
The dynamical nature of the self-energy allows to simultaneously describe both coherent and incoherent electronic excitations 
living on different energy scales, giving rise to a non-trivial renormalization of the spectral features. 
Real-space DMFT has been employed in the literature to describe many-body effects in a wide range of systems, 
lacking translational invariance in one or more spatial directions~\cite{snoekNJP10,jacobPRB82,valliPRL104,dasPRL107,valliPRB86,valliPRB91,valliPRB92,schuelerEPJST226,amaricciPRB95,amaricciPRB98,pudleinerPRB99,kropfPRB100,chioncelPRB92,droghettiPRB105,droghettiPRB106}, 
including graphene nanostructures~\cite{valliPRB94,valliNL18,valliPRB100,baumannPRA101,phungPRB102}. 
To address the transport properties, the transmission function is evaluated from the Landauer formula, 
where the Green's function is dressed with the many-body self-energy. 
This approach is a reasonable approximation 
when the system is not far away from equilibrium~\cite{meirPRL68,nessPRB82,jacobJPCM27,droghettiPRB95,droghettiPRB106} (see also SI).  

The local Coulomb repulsion between $\pi$-electrons was estimated by Parr {\it et al.}~\cite{parrJCP18} to be $U = \SI{16.93}{\eV}$, 
corresponding to $U/t \approx 6$, while recent studies suggest that a screened value,  
ranging between $U/t \approx 1.6$ for graphene and $U/t \approx 1.2$ for a benzene molecule, 
can supply for the missing long-range repulsion in the Hubbard model~\cite{schuelerPRL111}. 
In the following, we take $U$ as a parameter, which we vary within a reasonable range of values 
in order to explore the system behaviour from the weak- to the strong-coupling regimes. 
We perform zero-temperature calculations, which can be expected to be accurate as long as 
thermal excitation of charge carriers across the spectral gap are negligible,  
i.e., for $k_BT \ll \Delta_0$, where $k_B$ denotes Boltzmann's constant. 

\begin{figure}[t!]
\includegraphics[width=\linewidth, angle=0]{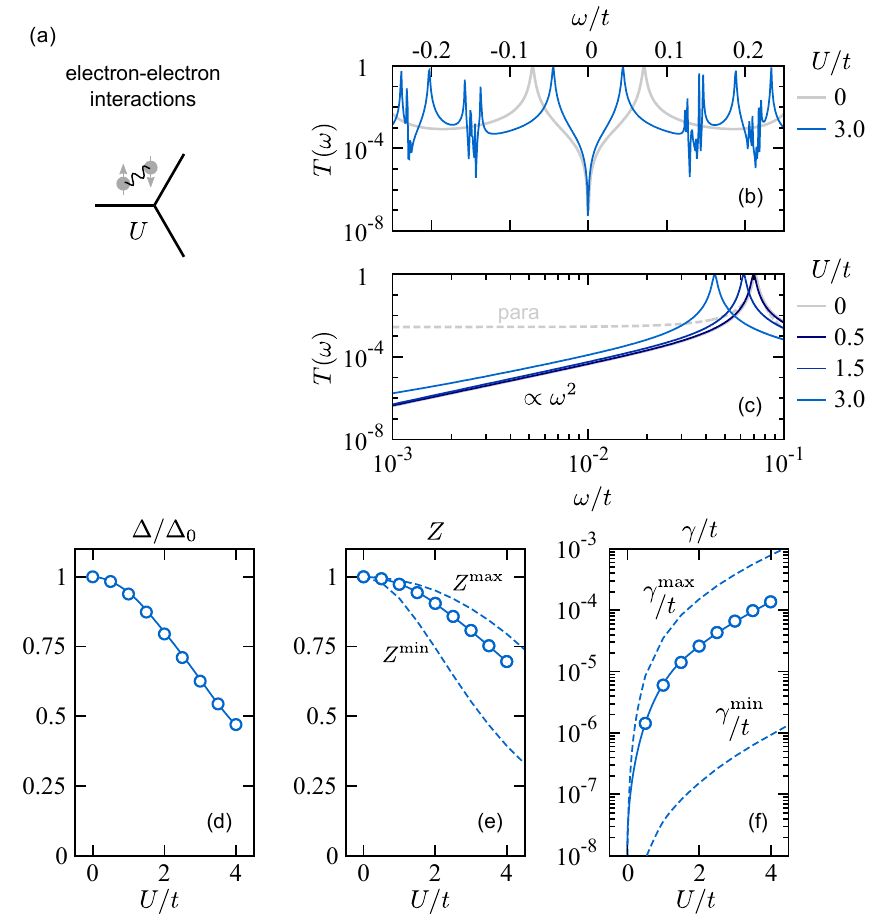}
\caption{(a) Schematic representation of the local Coulomb repulsion. 
(b) Many-body effects on the transmission function include a renormalization of the HOMO-LUMO gap, 
and a redistribution of spectral weight, giving rise to many-body resonances.  
(c) The universal behavior of the transmission function in the presence of the QI antiresonance
is preserved in the presence of electron-electron interactions. 
(d,e,f) Evolution of the spectral gap $\Delta$, of the average quasi-particle weight $Z$, and the average scattering rate $\gamma$ 
as a function of the Coulomb repulsion $U$. 
The renormalization of the gap $\Delta/\Delta_0$ correlates with the reduction of $Z$ and it is mostly controlled by its value at the edges $Z^{\mathrm{min}}$. 
The low scattering rate, i.e., $\gamma \ll \Gamma$, due to the lack of electronic states within the gap, 
cannot drive the electron transport away from the ballistic regime. }
\label{fig:effect_ee}  
\end{figure}

In Fig.~\ref{fig:effect_ee}(b,c) we show how many-body effects reflect on the transmission function of the meta configuration. 
The primary effect is a renormalization of the spectral gap $\Delta$ with respect to the tight-binding value $\Delta_0$, see also Fig.~\ref{fig:effect_ee}(d). 
While the low-energy structure is qualitatively identical to the one of the original tight-binding model, 
at higher energy scales, the system displays a significantly richer electronic structure, 
characterized by a redistribution of spectral weight between emergent many-body resonances.  
Remarkably, the universal $\omega^2$ behavior, as well as the QI antiresonance, 
survive in the presence of electron-electron interactions ranging from the weak- to the strong-coupling regime~\cite{valliNL18,valliPRB100}. 
The physics close to the Fermi energy, which is relevant for electron transport, 
can be rationalized ---to some extent--- in terms of key parameters derived from the many-body self-energy,
i.e., the quasi-particle weight $Z_{i}$ and scattering rate $\gamma_{i}$
\begin{eqnarray}
 Z _{i}           & = & \Big( 1 - \frac{\partial}{\partial\omega}\Re\Sigma_{ii}(\omega)\Big|_{\omega = 0} \Big)^{-1} \\
 \gamma_{i} & = & -Z_{i}\Im\Sigma_{ii}(0)
\end{eqnarray} 
which account for the renormalization of spectral features, and the (inverse) lifetime of electronic excitations, respectively. 
Since both depend on position, and hence display some degree of spatial distribution (see SI), 
it is also convenient to look at the spatial average, defined as ${\cal O} = n_C^{-1} \sum_i {\cal O}_i$ 
where ${\cal O}_i$ is a generic observable. 
In order to understand the behavior of those parameters as a function of the Coulomb repulsion, 
it is useful to recall that in the non-interactive limit ($U \rightarrow 0$) the many-body self-energy vanishes, therefore $Z \rightarrow 1$ and $\gamma \rightarrow 0$.

Within DMFT, the renormalization of the gap of a correlated insulator (or semiconductor) is controlled by the quasi-particle weight~\cite{sentefPRB80,valliPRB94,valliNL18,valliPRB100}. 
In a spatially-translational system, one would expect $\Delta = Z \Delta_0$. 
At the size considered here, the gap decreases \emph{faster} than the average $Z$ as a function of the Coulomb repulsion, 
and its behavior correlates well with the minimal value of $Z$ over the structure [$Z^{\mathrm{min}}$, compareFigs.~\ref{fig:effect_ee}(d,e)]. 
The lowest values of $Z$ correspond to edge C atoms, where the balance between potential and kinetic energy 
tilts more towards the former compared to bulk C atoms~\cite{valliPRB94,baumannPRA101}.  
The edges thus seem to control the reduction of the gap despite the average $Z$ being closer to the bulk values, see Fig.~\ref{fig:effect_ee}(e). 
Non-local correlations beyond real-space DMFT have been shown to \emph{enhance} the spectral gap~\cite{valliPRB91,pudleinerPRB99}. 
A possible interpretation is that the corrections arising from $\Sigma_{i\neq j}(\omega)$ renormalize the hopping parameters 
responsible for the splitting between the bonding (HOMO) and anti-bonding (LUMO) $\pi$ states.   
Taking into account such corrections is extremely challenging, and lies beyond the scope of the present analysis. 
However, note that our estimate of the spectral gap should be considered a \emph{conservative} approximation.

\begin{figure*}[thbp!]
\includegraphics[width=\linewidth, angle=0]{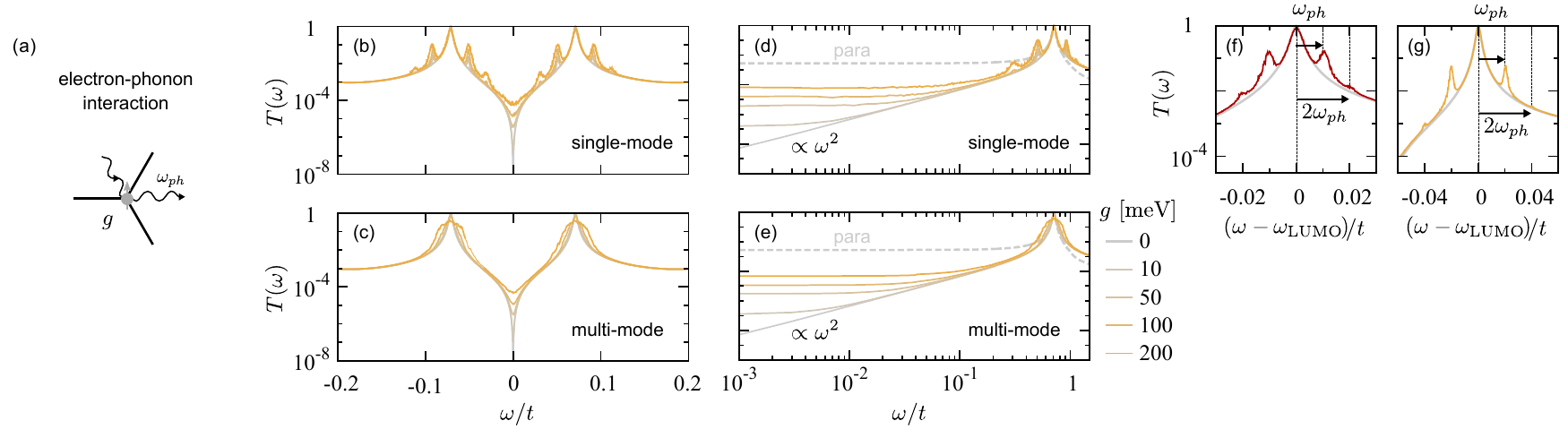}
\caption{(a) Schematic representation of the electron-phonon interaction. 
(b,c) Effects of the electron-phonon coupling on the transmission function. 
In the single-mode case, phonon satellite peaks are visible at integer multiples of $\omega_{ph}=\SI{26}{\milli\eV}$ around the frontier MOs. 
In the multi-mode case, the weighted average over $\omega_{ph}$ at $\SI{300}{\kelvin}$ broadens the resonances, 
and the transmission at the frontier MOs is no longer unitary.
(d,e) Evolution of the universal behavior of the transmission function across the weak- to the strong-electron-phonon coupling regimes. 
The transmission functions of the corresponding pristine para configurations is also given as a reference (dashed grey lines). 
(f,g) Single-mode transmission function around $\omega_{\mathrm{LUMO}}$ 
for $\omega_{ph}=\SI{13}{\milli\eV}$ (red line) and $\omega_{ph}=\SI{26}{\milli\eV}$ (orange line) and $g=\SI{100}{\milli\eV}$ compared to the result for $g=0$ (grey line). The arrows and the dashed lines highlight the phonon satellite peaks. }
\label{fig:effect_eph}  
\end{figure*}

The many-body scattering rate $\gamma$ is a measure of the dampening of the electronic excitations at the Fermi energy, 
while at finite energy, the self-energy induces broadening of the spectral and transport features. 
In a semiconductor (or insulator) the scattering rate is dramatically suppressed 
due to the lack of electronic states available for the scattering. 
As show in Fig.~\ref{fig:effect_ee}(f), $\gamma$ increases with the Coulomb repulsion, 
yet only becomes comparable with the interface scattering to the leads (i.e., $\gamma \sim \Gamma$) towards the strong-coupling regime. 
In this scenario, the scattering time $\tau_{\mathrm{e-e}} = \hbar/\gamma \approx \SI{100}{\femto\second}$ 
suggests that long-lived electronic excitations propagate through the graphene nanostructure. 
Assuming a typical value $v_F\approx \SI{1d6}{\metre \per \second}$ for the Fermi velocity in free-standing graphene~\cite{eliasNatPhys7}, 
the electron-electron scattering length can be estimated as $\lambda_{\mathrm{e-e}} = v_F \tau_{\mathrm{e-e}} \approx \SI{100}{\nano\metre}$, 
which is significantly longer than the size of the nanostructures considered here.  
This scenario is in stark contrast with Mott insulators, where the electronic excitation within the (correlated) energy gap 
features an extremely short lifetime~\cite{amaricciPRB85,waisPRB103} 
and molecules in a open-shell configuration~\cite{gandus2301.00282}.   

Hence, the primary effect of electronic-electron scattering is a renormalization of the spectral gap, 
while the electronic transport mechanism is dominated by coherent quasi-particles, 
and a QI antiresonance is clearly observed also in the presence of a strong Coulomb repulsion.

\subsection{Electron-phonon interaction}

Molecular vibrations are another possible source of scattering for electrons.  
In the literature, a strong inelastic vibrational signal has been reported close to electron transmission resonances through graphene nanoconstrictions~\cite{gunstPRB93},
and the phonon-induced dephasing has been suggested to be most relevant for molecular junctions in the meta configuration~\cite{tsujiJCP149}.
Indeed, evidence of a partial quenching of DQI in molecular junctions due to inelastic electron-phonon contributions 
has been reported both theoretically~\cite{markussenPRB89} and experimentally~\cite{rabacheJACS135,bessisSR6}. 
However, QI effects are not completely destroyed, not even at room temperature, 
as the conductance difference between junction in the meta and para configurations still remains substantial~\cite{markussenPRB89}. 

In the following, we analyze the effects of electron-phonon scattering on the electronic transmission through small graphene nanoflakes. Given the comparatively small sample size (compared to large-scale graphene devices) we are interested in coupling through bosonic vibrational modes rather than long-wavelength acoustic phonons \cite{BLYS2015,LBV2013}.
For the sake of simplicity, we consider the coupling to a dispersionless such mode, 
so that the coupled electron-phonon system is described by the Fr\"ohlich Hamiltonian 
\begin{equation}
 {\cal H}_{\mathrm{Fr\ddot{o}hlich}} = {\cal H}_0 + {\cal H}_{\mathrm{ph}} + {\cal H}_{\mathrm{e-ph}}
\end{equation}
with   
\begin{equation}\label{eq:Hph}
{\cal H}_{\mathrm{ph}} =  \omega_{ph} \Big(a^{\dagger} a + \frac{1}{2} \Big)
\end{equation}
and
\begin{equation}\label{eq:Heph}
{\cal H}_{\mathrm{e-ph}} =  g (a^{\dagger} + a) \sum_{i\sigma} n_{i\sigma}, 
\end{equation}
where $a$ ($a^{\dagger})$ is the bosonic annihilation (creation) operator of a phonon with energy $\omega_{ph}$, 
 coupled to the local electron density $n_i$ through a complex electron-phonon coupling matrix which is diagonal in the phonon subspace, 
$g=|g|\exp(\imath \phi)$, where $\phi$ is a random phase. 
We treat both $\omega_{ph}$ and $g$ as parameters, in order to understand the effect of electron-phonon scattering on the QI features in different regimes. 
For coupling to vibrational modes, prefactors $g = \lambda \omega_{ph}$ with $\lambda \approx 0.7$ have been estimated \cite{LBV2013}, corresponding to $g \approx \SI{20}{\milli\eV}$, i.e., at the lower end of the values we investigate here.
Typical values of the electron-phonon coupling for graphene on a substrate have been estimated to be $\SI{38}{\milli\eV}$ on SiO$_2$ and $\SI{62}{\milli\eV}$ on hBN~\cite{davydovPSS60}.

The expanded Hilbert space we simulate is spanned by a direct product  
of the electronic and phononic degrees of freedom, i.e., $n_C \cdot n_{ph}$, 
with $n_{ph}$ the number of phononic excitations. 
We compute the transmission $T_{0\rightarrow n}(\omega;\omega_{ph})$ with a fixed $\omega_{ph}$, 
from an incoming mode with $n=0$ phonons to each outgoing mode with $-n_{ph} \le n \le n_{ph}$ phonons. 
These contributions are added fully coherently 
by taking the trace in the Landauer formula over the phononic degrees of freedom. 
We also sample the phase $\phi$ of the electron-phonon coupling, 
in order to get the total transmission $T(\omega;\omega_{ph})$. 
We further extend our analysis from a single- to a multi-mode scenario by incoherently superimposing  
transmission functions $T(\omega;\omega_{ph})$ over a range of phonon frequencies, 
weighted with a Boltzmann factor evaluated at $T=\SI{300}{\kelvin}$ (i.e., $k_BT \approx \SI{26}{\milli\eV}$)
that takes into account the thermal occupation of the corresponding phonon mode 
(for a step-by-step description of all procedures see SI).

We calculate the transmission function for different values of the electron-phonon coupling $g$. 
In the single-mode approximation [Figs.~\ref{fig:effect_eph}(b,d)], 
we observe phonon satellite peaks around the frontier MO resonances, at integer multiples of the phonon frequency [Figs.~\ref{fig:effect_eph}(f,g)] for different values of $\omega_{ph}$. 
Focusing on the transmission function close to the antiresonance, 
we find that it is enhanced quadratically with the electron-phonon coupling, 
i.e., $T(\omega_{\mathrm{DQI}}) \propto g^2$,
 in close analogy to the effect of disorder. 
Continuing the analogy, we also observe that for a given value of the electron-phonon coupling, the transmission function interpolates 
between a constant regime at $\omega \approx \omega_{\mathrm{DQI}}$ 
and the characteristic $\omega^2$ behavior at higher energies. 
The crossover scale again increases with the electron-phonon coupling. 
Above a critical threshold, the transmission function becomes 
qualitatively indistinguishable from that of the para configuration.  
In the multi-mode case, instead of generating individual phonon satellites, 
the electron-phonon coupling broadens the resonances 
so that electron transport through the MOs is no longer unitary as shown in Figs.~\ref{fig:effect_eph}(c,e). 
The effects on the antiresonance are qualitatively similar to those observed in the single-mode scenario. 
The dichotomy between the effects close to the antiresonance and at higher energy scales 
appears to be a generic feature of the suppression of QI, which applies to different scattering mechanisms. 

Since in our numerical framework the electronic subsystem is no longer energy-conserving, 
nor we define an electronic self-energy,  
we cannot estimate a scattering length as we did in the case of disorder or electron-electron interactions.  
Therefore, we employ an alternative strategy that allows us 
to estimate the inelastic scattering length as a function of the electron-phonon coupling. 
In a nutshell, we consider the probability distribution 
of $T_{0\rightarrow n}(\omega=0;\omega_{ph})$ as a random walk in $n$. 
As the number of steps increases, we can fit a Gaussian distribution 
with width proportional to $\lambda_{\mathrm{e-ph}}/{\cal L}$ 
(the detailed procedure is described in the SI). 
For instance, we find  $\lambda_{\mathrm{e-ph}} \approx \SI{10}{\nano\meter}$ at $g=\SI{20}{\milli\eV}$ 
but we only reach the typical system size $\lambda_{\mathrm{e-ph}} \approx \SI{2}{\nano\meter}$ at $g=\SI{100}{\milli\eV}$, 
and a relation $\lambda_{\mathrm{e-ph}}(g) \propto g^{-1}$, analogous to the case of disorder.  
Upon comparison with the numerical data for the transmission function, 
this corresponds to the coupling range in which the characteristic $\omega^2$ behavior of the meta configuration 
is no longer observable and hence the effects of QI are contextually lost.   

In summary, phonons (or, more generally, inelastic scattering) limit DQI effects as soon as $k_B T$ exceeds the size of the HOMO-LUMO gap, in line with a much more simple estimate of smearing out the conductance on this energy scale. We therefore conclude that explicit consideration of electron-phonon scattering does not pose an additional limit to the observability of DQI.

\section{Discussion}\label{sec:conclusions}
The focus of the present work is the resilience of QI effects 
in the electron transport through graphene nanostructures. 
In the ballistic transport regime, the stability of DQI is rooted in its topological and symmetry origin, 
as also established in the recent literature. 
Furthermore, we investigated several effects, which can be possibly detrimental to QI, 
and for each we estimate a characteristic scale above which QI effects are likely to be lost. 

Our findings can be summarized as follows. 
(i) A size effects analysis reveals that in the ballistic regime, the in-gap transmission function 
displays a characteristic $\omega^2$ behavior that 
can be entirely ascribed to the presence of a QI antiresonance. 
In this regime, the transmission function is a universal function of the ratio 
between the molecule-lead coupling and the width of the gap, i.e., $\Gamma/\Delta_0$. 
In turn, it also determines an intrinsic resolution threshold necessary to resolve QI effects. 
(ii) The dominant effect of local many-body correlations due to the Coulomb repulsion 
is to renormalize the gap $\Delta < \Delta_0$
and, as long as thermal excitations of electrons across the gap is negligible, 
the electron-electron scattering rate is low and the effect on the QI properties is marginal. 
(iii) For disorder and electron-phonon scattering, 
we identify a similar behavior of the transmission function. 
Close to the antiresonance the transmission function is strongly enhanced 
with analogous scaling laws versus disorder strength $A$ and electron-phonon coupling $g$, 
whereas at higher energies the characteristic $\omega^2$ behavior is more resilient. 
For typical values reported in the literature, QI can be suppressed 
for graphene devices deposited on substrates like SiO$_2$ but not for cleaner devices deposited on hBN. 
The electron-phonon coupling in graphene (as well as in organic molecules) 
is typically low-enough that even at room temperature QI effects can survive.

While we investigated each case independently, multiple scattering sources are simultaneously in play. 
Hence, the effective electron lifetime is dominated by the process with the highest scattering rate 
(i.e., with the lowest lifetime or scattering length). 
Specifically, the overall scattering time is determined like a resistance in parallel
\begin{equation} 
 \frac{1}{\tau} = \frac{1}{\tau_{\Gamma}} + \frac{1}{\tau_{\mathrm{disorder}}} + \frac{1}{\tau_{\mathrm{e-e}}} + \frac{1}{\tau_{\mathrm{e-ph}}} + \ldots,  
\end{equation}
where the characteristic times correspond to processes involving electron scattering 
at the interface with the leads $\tau_\Gamma$, off disorder ($\tau_{\mathrm{disorder}}$), off other electronic ($\tau_{e-e}$) and phononic ($\tau_{e-ph}$) excitations, respectively, 
and similarly for any other possible process not explicitly included here.  

Within the present framework some effects can be combined with some additional effort, 
e.g., disorder with either electron-electron or electron-phonon interactions. 
For instance, we speculate that in the presence of the electronic correlations, 
the disorder-driven Fermi level alignment fluctuations could be reduced, 
especially for weak-to-moderate disorder $|\epsilon_i| \ll U$. 
The argument is that the Coulomb repulsion penalizes a spatially inhomogeneous charge distribution,~\cite{valliPRB94,baumannPRA101} 
and is expected to compete with disorder by renormalizing the disorder potential,
i.e., $\epsilon_i \rightarrow \epsilon_i + \Re\Sigma_{ii}(0)$ (see also SI). 
Other combinations, such as the simultaneous treatment of electron-electron and electron-phonon interactions
are very challenging, and are beyond the scope of the present analysis. 

In conclusion, our analysis provides a unified theoretical ground to explore the resilience of QI effects, 
and the necessary conditions for observing them under experimental conditions. 
At the same time, it emerges that the conductance alone may be insufficient 
for a characterization of the QI properties of a junction, 
while identifying non-linear $I-V_b$ characteristics can reveal 
the presence of an antiresonance even when partially concealed or suppressed 
due to the Fermi level alignment or electronic scattering.

\begin{acknowledgments}
We are thankful to A.~Amaricci and M.~Capone for useful discussion 
and for providing the L\'{a}nczos impurity solver.~\cite{weberPRB86,amaricciCPC273}  
AV and RS acknowledges financial support from the Austrian Science Fund (FWF) project number No.~P31631. 
TF and FL acknowledge support from FWF (DACH
proposal I3827-N36) and WWTF project MA14-002. 
Preliminary work for this project was also supported through the FWF Erwin Schr\"odinger fellowship J3890-N36. 
\end{acknowledgments}

\appendix
\section*{Methods}

\noindent \textbf{Electron transport.}
Within the Landauer formalism~\cite{landauerRD1} the electron transmission function is given by
\begin{equation} 
 T(\omega)=\mathrm{Tr}\Big[ \Gamma^{L}(\omega)G^{\dagger}(\omega)\Gamma^{R}(\omega)G(\omega)\Big], 
\end{equation}
where is $G$ is the retarded Green's function of the graphene nanostructure in the AO basis 
\begin{equation} \label{eq:Green}
 G(\omega) = \Big[\omega+\imath\eta - H -\Sigma^{L} -\Sigma^{R} \Big],
\end{equation}
and $\eta$ an infinitesimal.
The coupling to the leads is described in terms of the leads' self-energy, as 
\begin{equation}
 \Gamma^{\alpha}(\omega)=\frac{\imath}{2}\Big[\Sigma^{\alpha}(\omega)-\Sigma^{\alpha\dagger}(\omega)\Big],
\end{equation}
which is also a matrix in the AO basis. 
Within the WBL approximation the matrix elements $\Gamma^{L}_{\ell\ell}$ and $\Gamma^{R}_{rr}$ 
are given by an energy-independent constant $\Gamma$ for the edge C atoms $\{\ell\}$ and $\{r\}$, and zero otherwise. 
The applicability of the WBL approximation, in this context, is ensured by the observation that the spectral properties of the leads 
can modulate the transmission function and affect the Fermi level alignment 
but do not affect the existence of a QI antiresonance~\cite{verzijlJCP138,valliNL18,sengulNanoscale13}. \\

\noindent \textbf{Including disorder.}
In the presence of static disorder $\{\epsilon_k\}$, the transmission function $T(\omega; \{\epsilon_k\})$ is evaluated analogously 
as the pristine case, as the local energies enter the Green's function through the Hamiltonian.  
The disorder-averaged transmission is obtained as an \emph{incoherent} average over configurations of the random potential, i.e., 
\begin{equation} \label{eq:Te_disorder}
 T(\omega) = \sum_{\{\epsilon_k\}} T(\omega; \{\epsilon_k\}). 
\end{equation}
The transport is ballistic for each static configuration but not after the disorder averaging. \\

\noindent \textbf{Including electronic correlations.}
The effects of the electron-electron interactions are included in the Green's function (\ref{eq:Green}) though a many-body self-energy as 
\begin{equation} \label{eq:Green-many-body}
 G(\omega) = \Big[\omega+\imath\eta - H -\Sigma^{L} -\Sigma^{R} - \Sigma(\omega) \Big]. 
\end{equation}
For each frequency, the self-energy is a matrix in the AO basis obtained within a real-space DMFT approximation. 
In a nutshell, each locally inequivalent C atom of the nanostructure is mapped onto an auxiliary Anderson impurity model, 
which is solved by L\'{a}nczos exact diagonalization,~\cite{weberPRB86,amaricciCPC273} 
similarly as in previous works~\cite{valliPRB94,valliNL18,valliPRB100,baumannPRA101}.
The solution of the impurity problem yields a local self-energy $\Sigma_{ii}(\omega)$, so that the elements of the many-body self-energy 
are given by $\Sigma_{ij} (\omega)= \Sigma_{ii}(\omega)\delta_{ij}$, 
while elements $\Sigma_{i\neq j}$ are neglected~\cite{snoekNJP10,valliPRB86,valliPRB91}.
The procedure is iterated self-consistently, with an initial guess (typically zero) until convergence. \\

\noindent \textbf{Including electron-phonon coupling.}
Due to the interaction with phonons, it is necessary to consider an expanded Hilbert space spanned by a direct product  
of the electronic and phononic degrees of freedom, i.e., $n_C \cdot n_{\mathrm{ph}}$, with $n_{ph}$ the number of phononic excitations. 

For a given value of the phonon frequency $\omega_{\mathrm{ph}}$ the elements of the Fr\"ohlich Hamiltonian can be explicitly written as
\begin{equation} \label{eq:H0Heph}
 {\cal H} = \begin{pmatrix*}[l]
 \ddots	& 								&							&								& \\
		& {\cal H}_0+{\cal H}^{[-1]}_{\mathrm{ph}}	& {\cal H}_{\mathrm{e-ph}} 		&								& \\
		& {\cal H}_{\mathrm{e-ph}}^\dagger		& {\cal H}_0 					& {\cal H}_{\mathrm{e-ph}}			& \\
		&								& {\cal H}_{\mathrm{e-ph}}^\dagger	& {\cal H}_0+{\cal H}^{[1]}_{\mathrm{ph}}	& \\
		&								&							& 								& \ddots
 \end{pmatrix*},
\end{equation}
where ${\cal H}^{[n]}_\mathrm{ph}$ denoes the phonon Hamiltonian with $n$ phonons, ${\cal H}_\mathrm{e-ph}$ the electron-phonon coupling, 
and ${\cal H}_0$ the tight-binding Hamiltonian in the electron subspace. 

The electron-phonon coupling is a complex-valued matrix, 
which is diagonal in the phonon subspace, with elements $g=|g|\exp(\imath\phi)$. 
Hence, for each value $|g|$ we average the transmission function over the phases $\phi$ 
\begin{equation}
 T(\omega; \omega_{ph}) = \frac{1}{\pi} \int_0^{\pi}\mathrm{d}\phi \ T(\omega;\omega_{ph}, \phi),
\end{equation}
with
\begin{equation} \label{eq:landauer-eph}
 T(\omega;\omega_{ph}, \phi) = \mathrm{Tr} \Big[ \Gamma_L G_{\mathrm{e-ph}}^{\dagger}(\omega) \Gamma_R G_{\mathrm{e-ph}}(\omega) \Big].
\end{equation}
Note that Eq.~(\ref{eq:landauer-eph}) has the same form of the Landauer formula for the electronic system, 
but we denote with $G_{\mathrm{e-ph}}(\omega)=G(\omega;\omega_{ph},\phi)$ the Green's function in the extended Hilbert space, 
defined through Hamiltonian (\ref{eq:H0Heph}), and the $\Gamma_{\alpha}$ matrices are analogously extended, 
so that the trace includes both electronic and phononic degrees of freedom. 

Taking the trace corresponds to adding fully coherently all transmission contributions
$T_{0\rightarrow n}(\omega;\omega_{ph})$ for a fixed value of $\omega_{ph}$, 
from incoming modes with $n=0$ phonons to each outgoing mode with $-n_{ph} \le n \le n_{ph}$ phonons 
\begin{equation}
 T(\omega;\omega_{ph}, \phi) = \sum_n T_{0\rightarrow n}(\omega;\omega_{ph}, \phi). 
\end{equation}
In the weak-coupling regime, we verified that it is sufficient to restrict the phonon emission and absorption up to $n_{ph}=4$ phonon quanta. 

We extend our analysis of the transport properties from a single- to a multi-mode scenario for the phonons by incoherently superimposing  
transmission functions $T(\omega;\omega_{ph})$ over a range of phonon frequencies, 
weighted with a Boltzmann factor that takes into account the thermal occupation of the corresponding phonon mode 
\begin{equation}
 T(\omega) = \frac{1}{{\cal Z}}\int_{0}^{\Omega}\mathrm{d}\omega_{ph} T(\omega; \omega_{ph}) e^{-\beta\omega_{ph}}, 
\end{equation}
where the inverse temperature $\beta^{-1}=k_BT$ is evaluated at $T=\SI{300}{\kelvin}$, 
while $\Omega$ denotes an ultra-violet cutoff for the phonon excitation energy, 
and the partition function 
\begin{equation}
 {\cal Z} =\int_{0}^{\Omega}\mathrm{d}\omega_{ph} e^{-\beta\omega_{ph}} 
\end{equation}
ensures the proper normalization. 
Let us stress that the approximation behind the incoherent superposition of transmission functions calculated for different phonon energies 
is reasonable in the weak-coupling regime, where phonon-phonon scattering is unlikely.

\bibliographystyle{apsrev}
\bibliography{bibliography}


\pagebreak
\onecolumngrid

\setcounter{equation}{0}
\setcounter{figure}{0}
\setcounter{table}{0}
\setcounter{page}{1}
\makeatletter
\renewcommand{\thefigure}{S\arabic{figure}}
\renewcommand{\thepage}{S-\arabic{page}}

\newcommand*\mycommand[1]{\texttt{\emph{#1}}}

\begin{center}
  \textbf{\large Supporting Information: \\ Stability of destructive quantum interference antiresonance in electron transport through graphene nanostructures}\\[.2cm]
  A.~Valli,$^{1,2,*}$ T.~Fabian,$^{1}$ F.~Libisch,$^1$ and R.~Stadler$^1$\\[.1cm]
  {\itshape ${}^1$Institute for Theoretical Physics, Vienna University of Technology, Wiedner Hauptstrasse 8-10, A-1040 Vienna, Austria}\\
  {\itshape ${}^2$Department of Theoretical Physics, Institute of Physics, Budapest University of Technology and Economics, M\"uegyetem rkp. 3., H-1111 Budapest, Hungary}\\
\end{center}

\newpage

\section*{Remarks on quantum transport formalism}

\subsection*{Molecule-lead coupling and WBL approximation}

\noindent
The tight-binding Hamiltonian in the basis of a single $p_z$ atomic orbital per C atom reads
\begin{equation}\label{eq:tight-binding}
 {\cal H}_0 = t \sum_{\langle ij \rangle} \sum_{ \sigma} c^{\dagger}_{i\sigma} c^{\phantom{\dagger}}_{j\sigma}, 
\end{equation}
where $c^{(\dagger)}_{i\sigma}$ is the annihilation (creation) operator for an electron on site $i$ with spin $\sigma$, 
and $t$ is the hopping between nearest-neighbor sites $i$ and $j$. 
Within the Landauer formalism~\cite{landauerRD1}, the electron transmission function is given by
\begin{equation} \label{eq:landauer}
 T(\omega) = \textrm{Tr}\Big[ \Gamma^{L}(\omega)G^{\dagger}(\omega)\Gamma^{R}(\omega)G(\omega)\Big], 
\end{equation}
where the molecule retarded Green's function
\begin{equation}
 G(\omega) = \Big[ \omega + \imath\eta - {\cal H}_0 - \Sigma^L(\omega) - \Sigma^R(\omega) \Big]^{-1} 
\end{equation}
with $\eta \rightarrow 0^+$. 
The coupling to the leads are given in terms of the embedding self-energy 
\begin{equation}
 \Gamma^{\alpha}(\omega) = \frac{\imath}{2} \Big[ \Sigma^{\alpha}(\omega) - \Sigma^{\dagger \alpha}(\omega) \Big].
\end{equation}
In order to highlight the different contributions to the transmission function, the Landauer formula can be expressed also by expanding the trace explicit as
\begin{equation} 
 T(\omega) = \sum_{\ell \ell'} \sum_{rr'} \Gamma^{L}_{\ell \ell'}(\omega)G_{\ell' r'}^{\dagger}(\omega)\Gamma^{R}_{r' r}(\omega)G_{r \ell}(\omega), 
\end{equation}
where the sets $\{\ell, \ell'\}$ and $\{r, r'\}$ denote the AO connected to the left ($L$) and right ($R$) leads, respectively. 
Since the QI properties are determined by the topology of the molecular bridge rather than from the electronic structure of the leads, 
we make the following assumptions:
\begin{itemize}
 \item[(i)] that the molecule-lead coupling is diagonal in the AO basis,
 \item[(ii)] a wideband limit (WBL) approximation for the leads, 
 \item[(iii)] symmetric coupling between the AO of the molecule and the $L$ or $R$ leads.
\end{itemize}
The condition (i) implies that $\Sigma^{L}_{\ell\ell'}(\omega) = -\imath \Gamma_{\ell\ell}(\omega) \delta_{\ell\ell'}$, 
and analogously $\Sigma^{R}_{rr'}(\omega) = -\imath \Gamma_{rr}(\omega) \delta_{rr'}$, 
while within the the WBL aprroximation (ii) $\Sigma^{\alpha}(\omega) = -\imath \Gamma^{\alpha}$ is an energy-independent constant, 
and finally (iii) results in $\Gamma_{\ell\ell}=\Gamma_{rr}=\Gamma$.  
While the spectral properties of the leads modulate the transmission function, they do not directly affect the QI properties of the junction.~\cite{verzijlJCP138,valliNL18,sengulNanoscale13} 
Within the assumptions (i)-(iii) above, the transmission function takes a simplified form in terms of independent transmission channels $\ell \rightarrow r$ given by
\begin{equation}
 T(\omega) = \sum_{\ell r} T_{\ell \rightarrow r}(\omega) = \Gamma^2 \sum_{\ell r} |G_{\ell r}(\omega)|^2.
\end{equation}
Note that cross terms can arise if we relax the assumption of molecule-lead diagonal coupling. 
Elements $\Gamma_{\ell \neq \ell'}$ result in contributions to the transmission function of the form 
\begin{equation}
 T_{\ell,\ell' \rightarrow r}(\omega) = \Gamma_{\ell\ell'} \Gamma_{rr} |G_{\ell r}(\omega) + G_{\ell'r}(\omega)|^2,
\end{equation} 
thus inducing \textit{coherence} effects between transmission channels, 
which can, e.g., induce or suppress QI antiresonances within the HOMO-LUMO gap.~\cite{reuterJCP141,tsujiJCP141,hansenJPCC120,samangNJP19} 
However, we verified numerically that this is not the case for the systems we investigate here, 
and the general conclusions we draw remain valid also in the presence of coherent cross-channel contributions.

\subsection*{Transport through a disordered system}

\noindent
The local disorder is described by adding to the tight-binding Hamiltonian the term
\begin{equation}\label{eq:disorder}
 {\cal H}_{\mathrm{disorder}} = \sum_{i\sigma} \epsilon_i c^{\dagger}_{i\sigma}c^{\phantom{\dagger}}_{i\sigma}, 
\end{equation}
where $|\epsilon_{i}|$ are the randomly distributed on-site energies, with $\langle \varepsilon_i\rangle = 0$ and $A = \sqrt{\langle \varepsilon^2 \rangle}$ is the disorder strength. 
The transmission function $T(\omega; \{\epsilon_k\})$ for a specific realization of disorder is evaluated as in pristine case, through the Landauer formula (\ref{eq:landauer}). 
The disorder-averaged transmission is obtained as an \emph{incoherent} average over different realizations of the random potential, i.e., 
\begin{equation} \label{eq:Te_disorder}
 T(\omega) = \sum_{\{\epsilon_k\}} T(\omega; \{\epsilon_k\}). 
\end{equation}
The transport is ballistic for each static configuration but not after the disorder average.

\subsection*{Estimate of the disorder scattering length $\lambda_{\mathrm{disorder}}$ }

\begin{figure}[h]
\includegraphics[width=1.0\linewidth, angle=0]{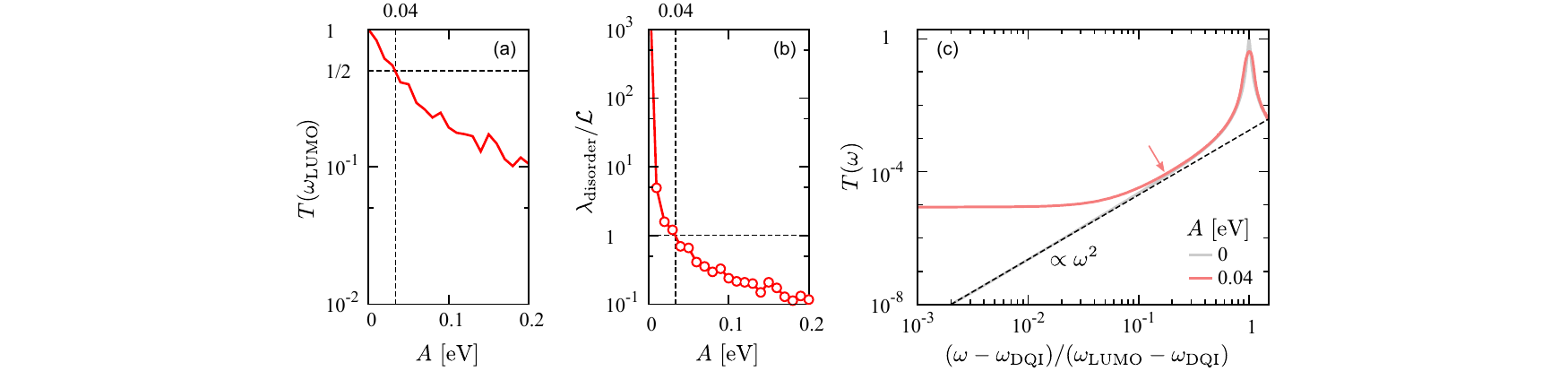}
\caption{(a) Reduction of the transmission function at the LUMO resonance for a $4 \times 4$ nanostructure, and 
(b) corresponding estimate of $\lambda_{\mathrm{disorder}}$ as function of disorder strength $A$. 
The dashed line indicate the critical value $A_{\mathrm{critical}}$, identified by the conditions $\lambda_{\mathrm{disorder}} \approx {\cal L}$ and $T=1/2$. 
(c) Transmission function for the pristine and the disordered system at $A \approx A_{\mathrm{critical}}$. }
\label{fig:lambda_disorder}  
\end{figure}

\noindent
Since the scattering off disorder is elastic, i.e., the electron energy does not change,  
we can obtain an estimate the scattering length as follows. 
We consider the transmission function for resonant scattering at, e.g., the HOMO or LUMO resonance. 
For any static disorder configuration the transport is ballistic, and therefore for a single open channel $T=1$, 
as we also verified in the numerical simulations.
Due to elastic scattering, the transmission is \textit{reduced} and given by~\cite{datta2005}
\begin{equation}
 T_{\mathrm{resonant}} = \frac{\lambda_{\mathrm{disorder}}}{{\cal L} + \lambda_{\mathrm{disorder}}}.
\end{equation}
Obviously, the ballistic limit is recovered for ${\cal L} \ll \lambda_{\mathrm{disorder}}$. 
It follows that (indicating $T_{\mathrm{resonant}}$ as $T$, for shortness)
\begin{equation}
 \frac{\lambda_{\mathrm{disorder}}}{{\cal L}} = \frac{T}{1-T}.
\end{equation}
For the $4 \times 4$ nanostructure, we can estimate ${\cal L} \approx \SI{1}{\nano\metre}$  
as the longest distance between the any pair ($\ell, r)$ of contact C atoms.  
In Fig.~\ref{fig:lambda_disorder} we show the reduction of the transmission function at the LUMO resonance, 
and well as the estimate of $\lambda_{\mathrm{disorder}}$ as a function of disorder strength $A$. 
We find, e.g., $\lambda_{\mathrm{disorder}} \approx \SI{1.2}{\nano\metre}$ at $A=\SI{30}{\milli\eV}$, 
with a dependence  $\lambda_{\mathrm{disorder}}(A) \propto A^{-1}$.  
Moreover, since one can expect QI effects to be lost for $\lambda_{\mathrm{disorder}} \approx {\cal L}$ (for which $T=1/2$), 
we can identify the critical value of the disorder strength $A_{\mathrm{critical}} \approx \SI{40}{\milli\eV}$ fulfilling this condition. 
For finite disorder, the depth of the DQI dip is reduced, and the transmission no longer follows $T\propto (\omega-\omega_{\mathrm{DQI}})^2$ behavior. 
Approaching the critical disorder strength, the DQI dip vanishes 
as qualitatively indicated by the arrow in Fig.~\ref{fig:lambda_disorder}(c). 

\subsection*{Transport in the presence of electron-electron interaction}

\noindent
To account for electron-electron interaction, we add to the tight-binding Hamiltonian a screened local Coulomb repulsion (i.e., a Hubbard interaction) 
described by  
\begin{equation}\label{eq:Hubbard}
 {\cal H}_{\mathrm{e-e}} = U \sum_i n_{i\uparrow} n_{i\downarrow} - \mu \sum_i (n_{i\uparrow} + n_{i\downarrow}),
\end{equation}
where the chemical potential $\mu=U/2$ ensures the Fermi energy to be at $\omega=0$ for any value of the Coulomb repulsion.  
The effects of electron-electron interactions are included in the Green's function though a many-body self-energy as 
\begin{equation} \label{eq:Green-many-body}
 G(\omega) = \Big[\omega+\imath\eta - {\cal H}_0 -\Sigma^{L}(\omega) - \Sigma^{R}(\omega) - \Sigma(\omega) \Big]^{-1}. 
\end{equation}
For each frequency, the self-energy is a matrix in the AO basis obtained within a real-space DMFT approximation. 
In a nutshell, each locally inequivalent C atom of the nanostructure is mapped onto an auxiliary Anderson impurity model, 
defined by the dynamical Weiss field
\begin{equation} \label{eq:AIM}
 {\cal G}^0_{i}(\omega) = \Big[ G_{ii}^{-1}(\omega) + \Sigma_{ii}(\omega) \Big]^{-1}
\end{equation}
with an initial guess for the many-body self-energy $\Sigma_{ii}(\omega)=0$. 
Each impurity model is solved by L\'{a}nczos exact diagonalization,~\cite{weberPRB86,amaricciCPC273} similarly to previous works~\cite{valliPRB94,valliNL18,valliPRB100,baumannPRA101}.
The solution of the impurity problem yields a local self-energy, so that the elements of the many-body self-energy 
are given by $\Sigma_{ij} (\omega)= \Sigma_{ii}(\omega)\delta_{ij}$, 
while elements $\Sigma_{i\neq j}(\omega)$ are neglected~\cite{snoekNJP10,valliPRB86,valliPRB91}.
As the many-body self-energy renormalizes the Green's function through (\ref{eq:Green-many-body}) 
and hence defines a corresponding impurity model (\ref{eq:AIM}),  
the procedure is iterated self-consistently until convergence. 

\noindent
Within this scheme, the electronic transmission function is evaluated through the Landauer formula (\ref{eq:landauer}), 
where the Green's function is renormalized by the many-body self-energy. 
In principle, the interactions renormalize not only the molecular Green's function but also the molecular-lead coupling, 
resulting in \textit{incoherent} contributions (which cannot be expressed in terms of an effective transmission function) 
and the electric current should be instead evaluated with the Meir-Wingreen formula.~\cite{meirPRL68}
However, the incoherent contributions are typically neglected in numerical simulations, 
which within a linear-response Kubo formalism for the current-current response function, 
corresponds to neglecting vertex corrections.~\cite{nessPRB82,jacobJPCM27,droghettiPRB95,droghettiPRB106}

\subsection*{Transport in the presence of electron-phonon interaction}

\noindent
Within the Landauer formalism, the electronic transmission function at a given energy $T(\omega)$ is evaluated from the Green's function $G(\omega)$. 
However, electron-phonon interaction changes the electron energy.
If we restrict ourselves to a single bosonic mode with fixed energy $\omega_{\mathrm{ph}}$, corresponding to lattice vibrations typically found
in the small graphene nanostructures considered here~\cite{LBV2013}, 
the possible electronic excitations become discrete and a combined treatment becomes numerically feasible.
We consider an expanded Hilbert space spanned by a direct product  
of the electronic and phononic degrees of freedom, i.e., $n_C \cdot n_{\mathrm{ph}}$, with $n_{ph}$ number of phononic excitations. 
The electron-phonon coupled system is described by a Fr\"ohlich Hamiltonian 
\begin{equation}
 {\cal H}_{\mathrm{Fr\ddot{o}hlich}} = {\cal H}_0 + {\cal H}_{\mathrm{ph}} + {\cal H}_{\mathrm{e-ph}}
\end{equation}
with   
\begin{equation}\label{eq:Hph}
{\cal H}_{\mathrm{ph}} =  \omega_{ph} \Big(a^{\dagger} a + \frac{1}{2} \Big)
\end{equation}
and
\begin{equation}\label{eq:Heph}
{\cal H}_{\mathrm{e-ph}} =  g (a^{\dagger} + a) \sum_{i\sigma} n_{i\sigma}, 
\end{equation}
where $a$ ($a^{\dagger})$ is the bosonic annihilation (creation) operator of a phonon with energy $\omega_{ph}$, 
which couples to the local electron density $n_i$ through a complex electron-phonon coupling matrix which is diagonal in the phonon subspace, 
$g=|g|\exp(\imath \phi)$, where $\phi$ is a random phase. 
The Fr\"ohlich Hamiltonian above can be explicitly as a matrix
\begin{equation} \label{eq:H0Heph}
 {\cal H} = \begin{pmatrix*}[l]
 \ddots	& 							&							&							& \\
		& {\cal H}_0+{\cal H}_{\mathrm{ph}}	& {\cal H}_{\mathrm{e-ph}} 		&							& \\
		& {\cal H}_{\mathrm{e-ph}}^\dagger	& {\cal H}_0 					& {\cal H}_{\mathrm{e-ph}}		& \\
		&							& {\cal H}_{\mathrm{e-ph}}^\dagger	& {\cal H}_0+{\cal H}_{\mathrm{ph}}	& \\
		&							&							& 							& \ddots
 \end{pmatrix*}
\end{equation}
The transmission function then is evaluated as 
\begin{equation} \label{eq:landauer-eph}
 T(\omega;\omega_{ph}, \phi) = \mathrm{Tr} \Big[ \Gamma^L(\omega) G_{\mathrm{e-ph}}^{\dagger}(\omega) \Gamma^R(\omega) G_{\mathrm{e-ph}}(\omega) \Big],
\end{equation}
which has the same form of the Landauer formula for the electronic system, 
but with $G_{\mathrm{e-ph}}(\omega)=G(\omega;\omega_{ph},\phi)$ the Green's function in the extended Hilbert space, 
defined through Hamiltonian (\ref{eq:H0Heph}), and the matrices $\Gamma^{L/R}$ are block diagonal,  
so that the trace includes both electronic and phononic degrees of freedom. 
Taking the trace corresponds to adding fully coherently all transmission contributions
$T_{0\rightarrow n}(\omega;\omega_{ph})$ with $-n_{ph} \le n \le n_{ph}$ phonons, at a given $\omega_{ph}$
\begin{equation}
 T(\omega;\omega_{ph}, \phi) = \sum_n T_{0\rightarrow n}(\omega;\omega_{ph}, \phi). 
\end{equation}
For a vanishing electron-phonon coupling $g \rightarrow 0$, only the mode with $n_{\mathrm{ph}}=0$ contributes to the transmission. 
At finite coupling, there are also contributions from channels $n_{\mathrm{ph}}>0$. 
In the weak-coupling regime, which is expected for reasonable values of the electron-phonon coupling, 
we verified that it is sufficient to retain absorption and emission processes with up to $n_{ph}=\pm 4$ phonon quanta. 
Finally, we average the transmission function over the phase $\phi$, as 
\begin{equation}
 T(\omega; \omega_{ph}) = \frac{1}{\pi} \int_0^{\pi}\mathrm{d}\phi \ T(\omega;\omega_{ph}, \phi). 
\end{equation}
We extend our analysis of the transport properties from a single- to a multi-mode phonon scenario by incoherently superimposing  
transmission functions $T(\omega;\omega_{ph})$ over a range of phonon frequencies, 
weighted with a Boltzmann factor that takes into account the thermal occupation of the corresponding phonon mode 
\begin{equation}
 T(\omega) = \frac{1}{{\cal Z}}\int_{0}^{\Omega}\mathrm{d}\omega_{ph} T(\omega; \omega_{ph}) e^{-\beta\omega_{ph}}, 
\end{equation}
where the inverse temperature $\beta^{-1}=k_BT$ is evaluated at $T=\SI{300}{\kelvin}$, 
while $\Omega$ denotes an ultra-violet cutoff for the phonon excitation energy, 
and the partition function 
\begin{equation}
 {\cal Z} =\int_{0}^{\Omega}\mathrm{d}\omega_{ph} e^{-\beta\omega_{ph}} 
\end{equation}
ensures proper normalization. 
The approximation behind the incoherent superposition of transmission functions calculated for different phonon energies 
is reasonable in the weak-coupling regime, where phonon-phonon scattering is unlikely. \\


\subsection*{Estimate of the electron-phonon scattering length $\lambda_{\mathrm{e-ph}}$ }

\begin{figure*}[h]
\includegraphics[width=1.0\linewidth, angle=0]{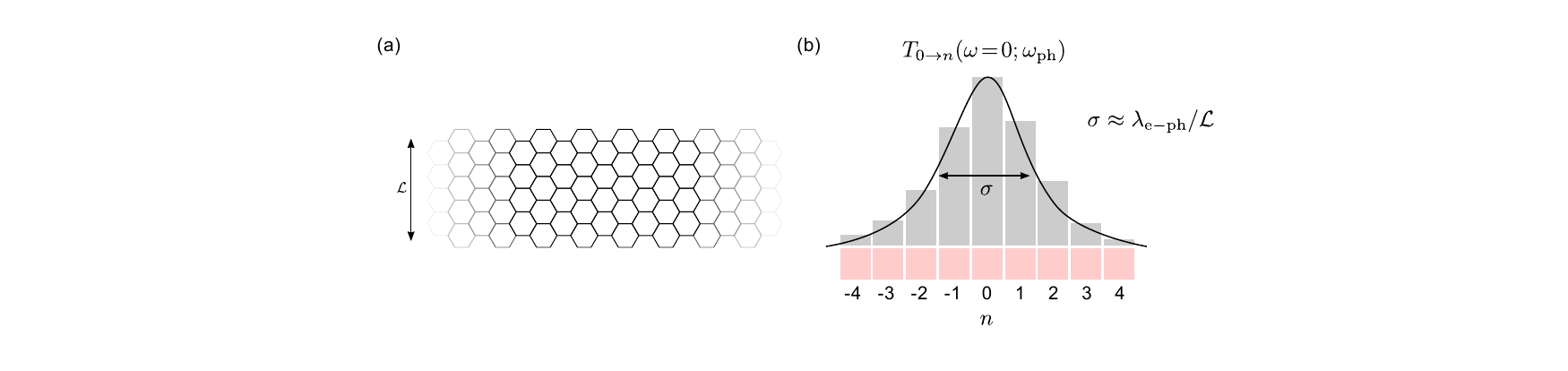}
\caption{(a) Auxiliary graphene nanoribbon with width ${\cal L}$. 
(b) Schematic histogram of the contributions $T_{n \rightarrow n}(\omega=0;\omega_{\mathrm{ph}})$ 
and a gaussian distribution fit (solid black line) to extract the variance $\sigma$ to estimate $\lambda_{\mathrm{e-ph}}$.  }
\label{fig:lambda_eph}
\end{figure*}

\noindent
Since the electron-phonon scattering is not elastic, 
we cannot follow the same strategy we used to estimate the scattering length in the presence of disorder, 
and we instead rely on an alternative procedure. 
The classical picture to our electron-phonon model is a random walk.  
An electron enters the lead with an initial energy $\varepsilon_0$ associated with an initial phononic occupation $n_0$, $\ket{\varepsilon_0}\otimes\ket{n_0}$. After a characteristic length scale $\lambda_{\mathrm{e-ph}}$, there will be, on average, one inelastic scattering event $\ket{\varepsilon_0}\otimes\ket{n_0}\rightarrow \ket{\varepsilon_0 \mp \omega}\otimes\ket{n_0 \pm 1}$. 
For $\mathcal L \gg \lambda_{\mathrm{e-ph}}$, the resulting distribution over levels with a final energy transfer of $n\omega$ into states $\ket{\varepsilon_0\mp n\omega}\otimes\ket{n_0\pm n}$ approaches a normal distribution centered around $n = 0$ (where $n$ counts the change in phononic occupation).  
By exploiting this analogy we obtain an estimate on the electron-phonon scattering length $\lambda_{\mathrm{e-ph}}(g)$. 
For a given coupling $g$, we compute the distribution of the transmission to the different phonon modes  
as a function of length ${\cal L}$ of an reference armchair graphene \textit{nanoribbon}, Fig.~\ref{fig:lambda_eph}(a), with the same width as the actual nanostructure.  
At ${\cal L} \ll \lambda_{\mathrm{e-ph}}$, the initial mode $\ket{\varepsilon_0}\otimes\ket{n_0}$ represents the dominant contribution to the transmission, 
while for increasing length also modes with $n\ne 0$ are populated. 
We fit the transmission (averaging over initial electron energy in a window of $\pm 0.01 \ \omega/t$ around $\omega=0$) for fixed $\omega_{\mathrm ph} = \SI{3}{\milli\eV}$ 
with a normal distribution ${\cal P}\{ T_{n \rightarrow 0}(\omega=0; \omega_{\mathrm{ph}}) \}$ (See Fig.~\ref{fig:lambda_eph}(b)).
If the variance of the distribution $\sigma = 1$, then we identify $\lambda_{\mathrm{e-ph}}$ with ${\cal L}$, 
while the distribution becomes broader (narrower), i.e., $\sigma > 1$ ($\sigma < 1$), for ${\cal L} > \lambda_{\mathrm{e-ph}}$ (${\cal L} < \lambda_{\mathrm{e-ph}}$).  

With this method, we can estimate $\lambda_{\mathrm{e-ph}} \approx \SI{1}{\micro\meter}$ for $g = \SI{0.2}{\milli\eV}$, 
and $\lambda_{\mathrm{e-ph}} \approx \SI{100}{\nano\meter}$ for $g = \SI{2}{\milli\eV}$, 
with a linear relation $\lambda_{\mathrm{e-ph}}(g) \propto g$. 
We validate the procedure against the numerical simulations, and we observe that 
for values of the electron-phonon coupling strength at which the estimated scattering length approaches the size of the nanostructure, 
the effects of DQI on the transmission functions are lost. \\

\newpage

\section*{Additional results}

\subsection*{Dependence of the transmission on $\Delta_0$ and $\Gamma$}

\begin{figure*}[h]
\includegraphics[width=1.0\linewidth, angle=0]{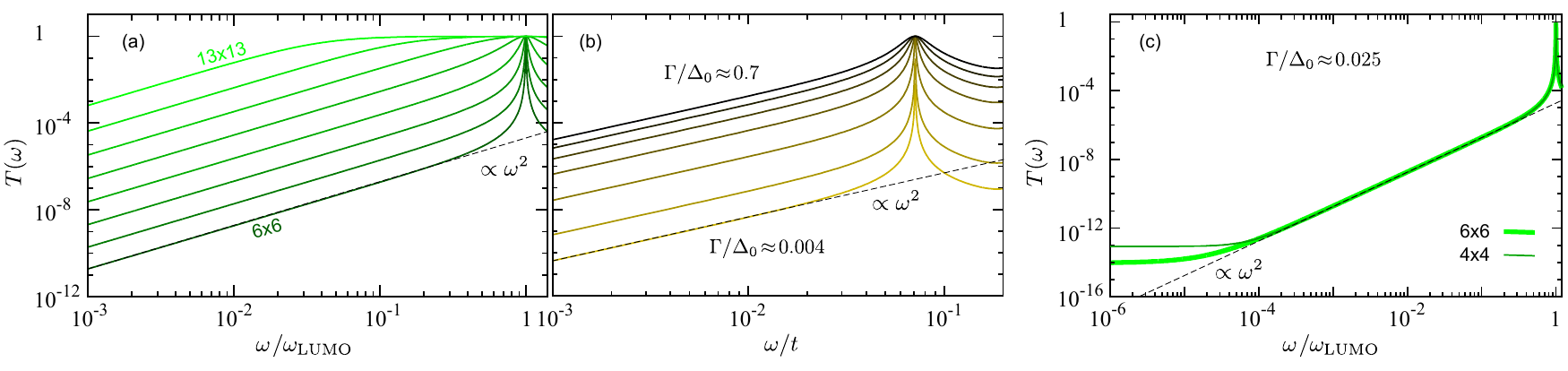}
\caption{Transmission function (a) through $N \times N$ nanostructure against $\omega/\omega_{\mathrm{LUMO}}$ for $\Gamma=0.0004t$,  
(b) through a $4 \times 4$ nanostructure for different values of the molecule-lead coupling $\Gamma$, and
(c) for nanostructures of different size but the same ratio $\Gamma/\Delta_0$. }
\label{fig:universal}
\end{figure*}

\noindent
Within the tight-binding approximation, besides the nearest-neighbor hopping $t$, 
the coupling to the leads $\Gamma$ and the HOMO-LUMO gap $\Delta_0$, are the other two energy scales of the problem. 
We have shown that in the meta configuration, the transmission function displays a characteristic behavior $\propto \omega^2$ within the gap, 
arising due to the existence of a QI antiresonance. 
In Figs.~\ref{fig:universal}(a,b), we show that the transmission functions has a similar dependence on $\Delta_0$ and on $\Gamma$. 
In order to have a better comparison, we align the LUMO resonances transmission function for different system size. 
In both cases, each curve corresponds to a different ratio $\Gamma/\Delta_0$. 
Instead, in Fig.~\ref{fig:universal}(c) we show two curves corresponding to nanostructures with different sizes, 
but with approximatively the same ratio $\Gamma/\Delta_0$. 
The data collapse in the whole energy window where $T(\omega) \propto \omega^2$ 
demonstrates that it is a \textit{universal} curve which depends only on the ratio $\Gamma/\Delta_0$. 
However, the saturation value as $\omega \rightarrow \omega_{\mathrm{DQI}}$ is not universal and depends on $\Gamma$.

\subsection*{Suppression of DQI: antiresonance}

\begin{figure*}[h]
\includegraphics[width=1.0\linewidth, angle=0]{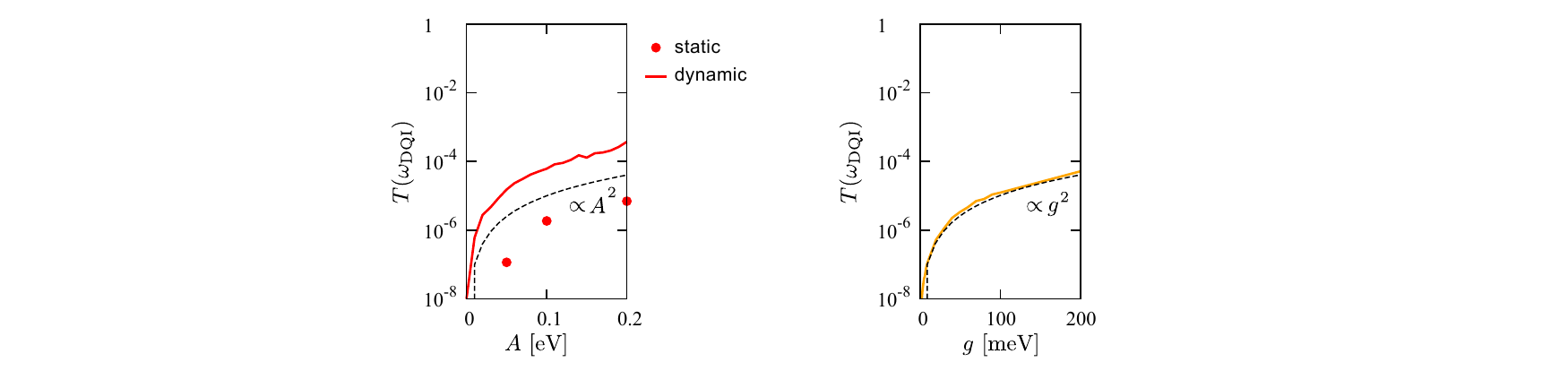}
\caption{Transmission function at the antiresonance $T(\omega_{\mathrm{DQI}})$ through a $4 \times 4$ nanostructure, 
versus (a) disorder strength and (b) electron-phonon coupling. 
QI effects are suppressed, as $T(\omega_{\mathrm{DQI}})$ is enhanced quadratically as a function of both control parameters. }
\label{fig:qi_suppression}
\end{figure*}

\noindent
One of the most evident effects of disorder and of electron-phonon scattering is the enhancement of the transmission function 
close to the antiresonance, as DQI is progressively lost, as shown in Fig.~2(b,c,d,e) in the manuscript.
We analyze the dependence of $T(\omega_{\mathrm{DQI}})$ in both scenarios to extract its dependence on the control parameter.  
In Fig.~\ref{fig:qi_suppression}(a) we show that $T(\omega_{\mathrm{DQI}}) \propto A^2$ as a function of disorder strength for both static and dynamic disorder.  
The main difference, is that the enhancement is significantly stronger (about two orders of magnitude) after averaging over the disorder realizations.  
In Fig.~\ref{fig:qi_suppression}(b) we show that that $T(\omega_{\mathrm{DQI}}) \propto g^2$ as a function of the electron-phonon coupling, 
and the enhancement appears to be intermediate between those of static and dynamic disorder.

\newpage

\subsection*{Suppression of DQI: $I-V_b$ characteristics}

\noindent
Within our numerical framework, the electron current (per spin) is evaluated from the transmission function as 
\begin{equation}
 I = \frac{e}{h} \int_{-\infty}^{\infty} \mathrm{d}\omega \ T(\omega; eV_b) \Big[ f\Big(\omega-\frac{eV_b}{2}\Big) - f\Big(\omega+\frac{eV_b}{2}\Big)\Big],
\end{equation}
with the Fermi distribution function for the electrodes given by 
\begin{equation}
 f(\omega) = \frac{1}{1+\exp(\omega/k_B T)} 
\end{equation}
where $e$ denotes the electric charge, $h$ the Planck constant, $k_B$ the Boltzmann constant, 
and $V_b$ is the symmetric bias drop between the source and the drain. 
The characteristic behavior of the transmission function reflects on the $I-V$ characteristics. 
This is easiest to evaluate assuming $T(\omega,V_b) \approx T(\omega)$, 
and $\omega \ll k_BT$ so that the Fermi distribution function can be approximated with its derivative, 
hence restricting the energy integral within the bias window
\begin{equation}
 I \approx \frac{e}{h} \int_{-eV_b/2}^{eV_b/2} \mathrm{d}\omega \ T(\omega).
\end{equation}
Therefore, we can identify two transport regimes, which can be identified far-enough from the resonant transport condition, i.e., $|\omega| \lesssim |\omega_{\mathrm{LUMO}}-\Gamma|$. 
In the para configuration, $T(\omega) \sim \mathrm{const.}$ within the HOMO-LUMO gap, and the integral yields $I \propto V_b$. 
In the meta configuration, the transmission saturates close to the antiresonance, i.e., $T(\omega) \sim \mathrm{const.}$ for $\omega \approx \omega_{\mathrm{DQI}}$ 
but at higher energies there is a crossover to a non-linear transport regime, 
as $T(\omega) \propto \omega^2$ and the integral yields $I \propto V_b^3$. 

\begin{figure}[h]
\includegraphics[width=1.0\linewidth, angle=0]{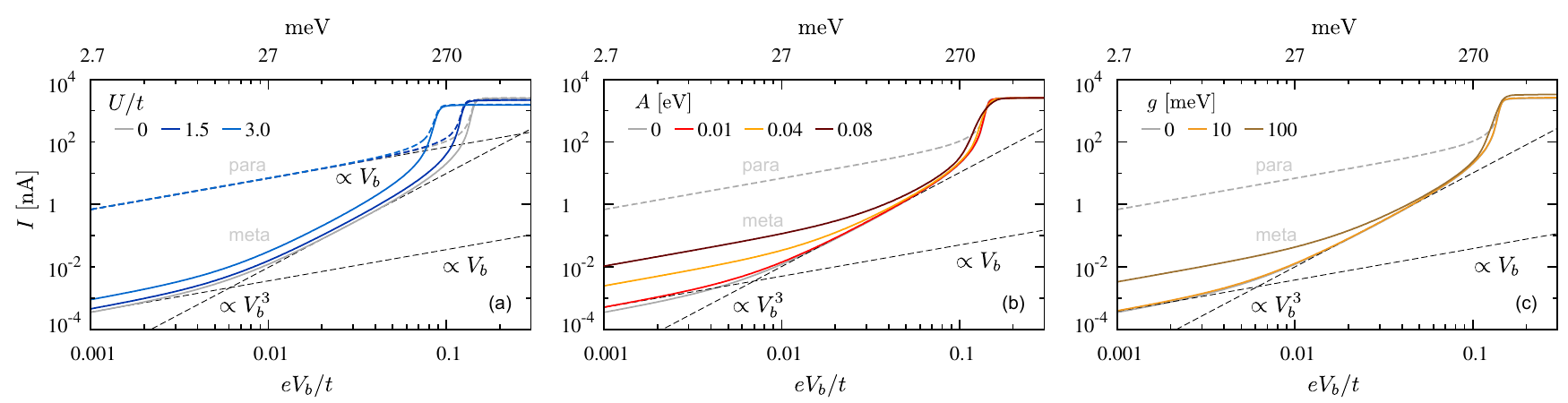}
\caption{Typical behavior of the I-V characteristics through $4 \times 4$ nanostructure in the meta (solid lines) and para (dashed lines) configurations 
in the presence of (a) electron-electron interactions, (b) dynamical disorder, and (c) electron-phonon scattering. 
The characteristic linear $I \propto V_b$ and non-linear $I \propto V_b^3$ regimes are indicated by dashed lines, as labelled. }
\label{fig:IV_behavior}  
\end{figure}

\noindent
The two transport regimes above can be clearly identified in the numerical simulations, as shown in Fig.~\ref{fig:IV_behavior}. 
Specifically, Fig.~\ref{fig:IV_behavior}(a) shows the $I-V_b$ characteristics for different values of the local Coulomb repulsion $U$. 
The net effect of electron-electron interaction is to renormalize the gap, as can be observed from the shift of the current plateau to lower bias voltages, 
which also results in a weak enhancement of the current at low $V_b$. 
However, neither the linear regime in the para configuration, nor the crossover behavior from the linear to the non-linear regime in the meta configuration, 
are qualitatively modified upon increasing $U$. 
The scenario is different in the case of disorder and electron-phonon coupling, which are shown in Figs.~\ref{fig:IV_behavior}(b,c), respectively.  
The suppression of QI manifest in (i) an enhancement of the current at low $V_b$, 
and (ii) an extension of the linear regime to higher $V_b$. 
However, the non-linear regime is not destroyed until critical values of the control parameters, i.e., the disorder strength or the electron-phonon coupling. 
The critical disorder strength is $A_{\mathrm{critical}} \approx \SI{40}{\milli\eV}$, 
in agreement with the estimates obtained from the analysis of the scattering length $\lambda_{\mathrm{disorder}}$.

\newpage

\subsection*{Energy distribution of Green's function zeros}

\begin{figure}[h]
\includegraphics[width=1.0\linewidth, angle=0]{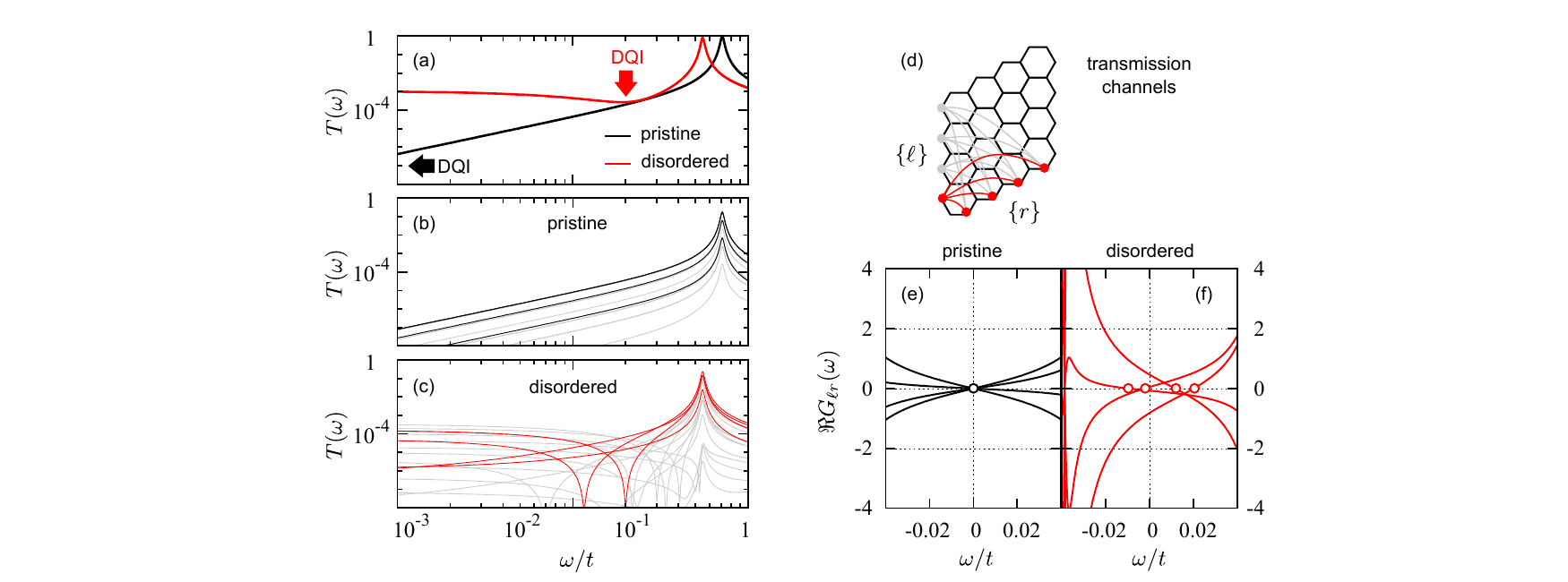}
\caption{(a) Transmission function through the pristine $4 \times 4$ nanostructure and for a specific disorder configuration, as labelled. 
(b,c) Individual transmission channel contributions. The color (grey) lines correspond to the channels $\ell \rightarrow r$ shown in panel (d). 
(e,f) Real part of the Green's function $\Re G_{\ell r}(\omega)$ for pristine (e) and disordered (f) configurations 
for selected channels, marked in color in panel (d). 
In the pristine system, all nodes (empty black circles) are at $\omega_{\mathrm{DQI}}=0$, resulting in a narrow antiresonance, 
while disorder results in a distribution of nodes (empty red cicles) at different energies, responsible for a partial cancellation of the transmission.}
\label{fig:wdqi_disorder}  
\end{figure}

\noindent
The main effect of static disorder is to shift and broaden the QI antiresonance, see Fig.~\ref{fig:wdqi_disorder}(a).  
We shed some light on the mechanism behind this effect. 
For diagonal molecule-lead coupling and the WBL approximation, the condition for DQI in a given channel $\ell \rightarrow r$ 
is then connected to the existence of a zero of a specific Green's function component $G_{\ell r}(\omega)$. 
Since $\Im G_{\ell r}(\omega) \approx 0$ 
far away from resonant transport, 
i.e., for $|\omega - \omega_{k-\mathrm{MO}}| \gg \Gamma$, 
where $\omega_{k-\mathrm{MO}}$ is the energy of the $k$-th MO, 
the condition then reduces to 
\begin{equation} \label{eq:reg0}
 \Re G_{\ell r}(\omega_{\mathrm{DQI}}) = 0. 
\end{equation}
In Figs.~\ref{fig:wdqi_disorder}(b,c) we plot the individual contributions of all transmission channels $\{\ell\} \rightarrow \{r\}$. 
The curves in color (grey) solid lines correspond to the channels marked by color (grey) connectors in Fig.~\ref{fig:wdqi_disorder}(d). 
In the pristine system, despite the individual contributions to the transmission function are different, 
all channels display a node in $\omega_{\mathrm{DQI}}=0$. 
In the presence of disorder there is a distribution of nodes at different energies. 
This is shown explicitly by plotting $\Re G_{\ell r}(\omega)$, as shown in Figs.~\ref{fig:wdqi_disorder}(e,f) 
for the selected transmission channels marked in color in Fig.~\ref{fig:wdqi_disorder}(d). 

\end{document}